\documentclass[prd, aps, nofootinbib, preprintnumbers,
showpacs,superscriptaddress,twocolumn]{revtex4}

\usepackage{graphics,graphicx,rotating}
\usepackage[usenames]{color}
\usepackage[percent]{overpic}
\usepackage{mathrsfs,amsmath,amssymb}
\usepackage{wasysym}
\usepackage{times}
\usepackage{mathptmx}

\newcommand{\be}{\begin{equation}}
\newcommand{\ee}{\end{equation}}
\newcommand{\beq}{\begin{equation}}
\newcommand{\eeq}{\end{equation}}
\newcommand{\ber}{\begin{eqnarray}}
\newcommand{\eer}{\end{eqnarray}}
\newcommand{\bea}{\begin{eqnarray}}
\newcommand{\eea}{\end{eqnarray}}

\newcommand{\AEI}{\affiliation{Max-Planck-Institut f\"ur
    Gravitationsphysik, Am M\"uhlenberg 1, 14475 Potsdam, Germany}}

\newcommand{\palma}{\affiliation{Departament de F\'isica, 
  Universitat de les Illes Balears, Cra.\ Valldemossa Km.\ 7.5, Palma 
de Mallorca, E-07122 Spain}}

\newcommand{\vienna}{\affiliation{Faculty of Physics, University of Vienna, Boltzmanngasse 5, A-1090 Vienna, Austria}}

\newcommand{\LIGOCaltech}{\affiliation{LIGO Laboratory, California Institute of Technology, 
Pasadena, CA 91125, USA}}
\newcommand{\TAPIR}{\affiliation{Theoretical Astrophysics, California Institute of Technology, 
Pasadena, CA 91125, USA}}

\begin{document}

\title{Length requirements for numerical-relativity waveforms}

\author{Mark Hannam}        \vienna
\author{Sascha Husa}        \palma
\author{Frank Ohme}         \AEI
\author{P. Ajith}           \LIGOCaltech \TAPIR

\date{\today}

\begin{abstract}
One way to produce complete inspiral-merger-ringdown gravitational waveforms from
black-hole-binary systems is to connect post-Newtonian (PN) and numerical-relativity (NR)
results to create ``hybrid'' waveforms. Hybrid waveforms are central to the construction of 
some phenomenological models for GW search templates, and for tests of GW 
search pipelines. The dominant error source in hybrid waveforms arises from the PN contribution, 
and can be reduced by increasing the number of NR GW cycles that are included in the hybrid. 
Hybrid waveforms are considered sufficiently accurate for GW detection if their mismatch
error is below 3\% (i.e., a fitting factor above 0.97). We address the 
question of the length requirements of NR waveforms such that the final hybrid waveforms 
meet this requirement, considering nonspinning binaries with 
$q = M_2/M_1 \in [1,4]$ and equal-mass binaries with $\chi = S_i/M_i^2 \in [-0.5,0.5]$.
We conclude that for the cases we study simulations must contain between three 
(in the equal-mass nonspinning case) and ten (the $\chi = 0.5$ case) orbits before merger,
but there is also evidence that these are the regions of parameter space for which the
\emph{least} number of cycles will be needed.  
\end{abstract}


\maketitle

\section{Introduction}

Numerical simulations play a key role in efforts to detect gravitational waves 
(GWs) from compact-binary coalescences, in particular from black-hole binaries. 
One of their most direct applications for GW searches has been as input in 
phenomenological~\cite{Ajith:2007qp,Ajith:2007kx,Ajith:2007xh,Ajith:2009bn,Santamaria:2010yb,Sturani:2010yv} 
and effective-one-body 
(EOB)~\cite{Buonanno:2007pf,Damour:2007yf,Damour:2007vq,Damour:2008te,Baker:2008mj,Mroue:2008fu,Damour:2009kr,Buonanno:2009qa,Pan:2009wj} 
waveform models, which can in turn be used
to construct template banks of theoretical waveforms for use in 
matched-filter searches in detector data. As part of the NINJA project~\cite{ninja-wiki}
they are also used to test a battery of search pipelines: the numerical-relativity
(NR) waveforms are injected into simulated detector noise and the search
pipelines attempt to find them~\cite{Aylott:2009tn,Aylott:2009ya}. 

Waveform models are constructed by combining information from 
post-Newtonian (PN) calculations of the long slow inspiral of the 
binary~\cite{lrr-2006-4}, 
and NR simulations of the last orbits and merger. The PN approximation 
becomes less accurate as the binary approaches merger, and we will
argue in this paper that at the current state of the art, PN errors dominate 
significantly over errors in the NR waveforms. 
Therefore more accurate waveform models can be produced 
by using less PN inspiral cycles, and correspondingly more  NR cycles.
The question we begin to address in this paper is: how many NR cycles 
are necessary in order to
produce waveform models that are sufficiently accurate for GW detection?
By this we mean that the mismatch error in the full waveform (minimized
with respect to all search parameters) is less than 
3\%, as described in Sec.~\ref{sec:definitions}. 

This question is important because NR waveforms from an extremely large 
number of binary configurations may be necessary to produce waveform
models that accurately represent the entire black-hole-binary parameter space. 
NR simulations are computationally expensive, and the computational cost
grows drastically with the length of the simulation. If the number of inspiral
cycles in a simulation is to be increased by a factor of two, for example,
the simulation doesn't only have to run for twice as long. To accurately 
capture the phase evolution of the binary over this extended time, higher
numerical resolutions are required, and the overall increase in the 
cost of the simulation may be by factors of ten, both in memory usage
and the time the simulation takes to run. 

Current ``long'' simulations cover $\sim$10 GW cycles before 
merger~\cite{Hannam:2009rd},
and the longest simulation to date covers about 30 
cycles~\cite{Boyle:2007ft}. The task of exploring the full black-hole-binary 
parameter space with long numerical simulations is a large-scale computational
challenge that will take several years with current methods. On such a time
scale, it is crucial to understand as much as possible about any requirements
that may affect the overall cost by an order of magnitude!

The necessary length of NR waveforms will change with improvements
in the accuracy of PN methods and increased understanding of EOB methods. 
As such, we can only make a first step in considering waveform length 
requirements. Our method is to study \emph{hybrid} waveforms, which are a 
simple connection of PN and NR results, and allow us to focus on the
effects of the PN errors without having to take into account any additional
artifacts that might arise in a given waveform-model construction procedure.
We believe that this allows us to make the most conclusive statements that
are possible with the current state of the art. 

We summarize mismatch calculations in Sec.~\ref{sec:definitions}, 
describe our numerical waveforms in Sec.~\ref{sec:waveforms} and the production
of hybrid waveforms in Sec.~\ref{sec:hybrids}. We use hybrids based on two
time-domain PN approximants, the TaylorT1 and TaylorT4 approximants, the accuracy 
of which we have already studied over a 10-cycle range before merger 
in~\cite{Hannam:2010ec}. 
We put our previous PN-NR comparisons into the context of GW searches
by computing the mismatch between hybrids with differing numbers of TaylorT1 and NR
cycles in Sec.~\ref{sec:T1hybrids}, and demonstrate that the hybrid errors are
indeed dominated by PN effects. We then turn to the question of length requirements
for NR waveforms. 

We will explain and justify our approach in Sec.~\ref{sec:length}. Since it has
already been shown that pure PN waveforms truncated before merger are
sufficient to detect nonspinning binaries with masses up to 
$12\,M_\odot$~\cite{Buonanno:2009zt}, we consider binaries with total mass $M >
10\,M_\odot$, 
and our analysis is with respect to the Advanced LIGO detector~\cite{AdLIGO}. 
Our study covers a significant portion of the parameter
space that has already been treated by NR simulations: nonspinning binaries with
mass ratios up to $q = M_2/M_1 = 4$, and equal-mass binaries with equal, 
non-precessing spins up to $\chi = \pm 0.5$. \
Results for
nonspinning binaries are given in Sec.~\ref{sec:lengthUM}, and spinning binaries are
treated in Sec.~\ref{sec:lengthSpin}. We will not consider phenomenological or 
EOB methods, and we will only 
briefly touch on parameter estimation in Sec.~\ref{sec:PE}. Our first
conclusions about waveform length requirements are summarized and discussed
in the Conclusion.

\section{Mismatches and GW detection}
\label{sec:definitions}

The quantity that we will use to assess the detectability of our hybrid waveforms is the 
\emph{mismatch}. The standard requirement for current GW searches is that the members of
a template bank be separated such that the no more than 10\% of signals would be lost
due to the mismatch between the signal and the template, 
and this translates into a mismatch between neighboring templates of less than 
3\%~\cite{Abbott:2009qj,Abbott:2009tt}.
However, one must also take into account
the mismatch due to errors in the theoretical waveforms; ultimately we want the 
\emph{sum} of the template-spacing mismatch and the waveform-error mismatch to be less than
3\%. We are then free to decide how to divide that mismatch between template spacing
and waveform error. For example, the authors of the waveform accuracy study 
in~\cite{Lindblom:2008cm} 
request that the waveform-error mismatch be below 0.5\%. We could alternatively argue 
that the waveform error can be close to 3\%, and the template spacing should be made 
correspondingly small. This question deserves further attention, but in this work we will
consider 3\% as the basic requirement, and also quote results consistent with a maximum 
mismatch requirement of 1.5\% and 0.5\%. 

For two GW signals $h_1$ and $h_2$, we define an 
inner product in the Fourier domain weighted by the power spectral density of
the detector noise, $S_n(f)$,
as~\cite{Cutler94},
 \beq
   \label{eq:scalar_prod}
 \langle h_1| h_2\rangle := 4 \, {\rm Re} \left[ \int_{f_{\rm
         min}}^{f_{\rm max}} 
     \frac{ \tilde h_1(f) 
     \tilde h_2^\ast (f)}{S_n(f)} \, df \right] \, ,
 \eeq  
 where $[f_{\rm min},f_{\rm max}]$ is the intersection of the chosen sensitivity 
 range of the detector ($[20,10^4]$\,Hz in this study) and the range of validity
of the waveform data (in most cases $fM \in [1.25\times10^{-3},0.15]$).
 $\tilde h_1(f)$ and  $\tilde h_2(f)$ denote the Fourier transform of $h_1$ and
$h_2$, respectively. 
 Our data in general represent the Weyl scalar $\Psi_4(t)$, not the wave strain
$h(t)$, but the two 
 are related by $\Psi_4 = \ddot h_+ - i \ddot h_\times$, and two time
integrations 
 can be performed trivially in the frequency domain. 
  
Given the definition of the inner product $\langle h_1| h_2\rangle$, we normalize it
and maximize over phase and time offsets in the data. 
This is the faithfulness of the waveform: it is a measure
of how ``far'' a theoretical waveform is from a supposedly true waveform with the same
physical parameters. The \emph{faithfulness mismatch} is the deviation of the faithfulness
from unity: \beq
{\cal M} = 1 -  \max_{\tau,\Phi} \frac{\langle h_1 | h_2 \rangle }{\sqrt{
     \langle h_1|h_1\rangle \langle h_2|h_2\rangle }}. 
\eeq

In a GW search, the goal is essentially to find the template-bank member that has the smallest 
mismatch with the detector data. We are therefore really interested in
the mismatch optimized over all of the members of a theoretical waveform family, i.e., 
minimized with respect to the binary's intrinsic physical parameters (the mass, mass ratio 
and spins), and the extrinsic parameters (the position of the
binary in the sky and its orientation). The fully optimized mismatch is the same as
$(1-F)$, where $F$ is the waveform's \emph{fitting factor}. 
In our study we have access to signals of
isolated binary configurations, and the only parameter we can optimize with respect to 
is the total mass. As such, an optimization of the mismatch with respect to total mass
can be no more than an upper bound on the full optimized mismatch. But we will argue
in Sec.~\ref{sec:lengthUM} that optimization with respect to the other physical parameters 
will not qualitatively
alter our results. In Sec.~\ref{sec:hybrids} and \ref{sec:T1hybrids}, where we assess the 
sources of physical error in our hybrids, we will not optimize with respect to total mass. 
But in Secs.~\ref{sec:length} -- \ref{sec:PE}, where we make calculations relevant to 
GW detection, we will use the mass optimization, and intend these results to be a 
reasonable estimate of the mismatch calculated from the fully optimized fitting factor. 

Our main focus in this work is GW detection, but we will also refer to a quantity that 
is relevant to parameter estimation. If we have two theoretical waveforms for the same 
physical system, $h_1$ and $h_2$, each with some associated uncertainty, then we can 
define $\delta h = h_1 - h_2$ and then $|| \delta h||^2 = \langle \delta h |
\delta h \rangle$. 
If $|| \delta h|| < 1$, then the two waveforms would be indistinguishable in a detector 
measurement~\cite{Lindblom:2008cm}. The indistinguishability of two waveforms 
depends on the signal-to-noise ratio (SNR) of the detection: if the SNR is
sufficiently 
low, then any two waveforms will be indistinguishable (although below an SNR of 
eight they will not be considered detectable anyway), and for two waveforms of
arbitrarily low $||\delta h||$, we can always find an SNR high enough such that they can 
in fact be distinguished. For example, in~\cite{Hannam:2009hh} we found that current NR 
equal-mass nonspinning waveforms from five different codes would be indistinguishable 
in the Advanced ground-based detectors if the SNR is below 25. And 
in~\cite{Santamaria:2010yb} it was shown that $q=2$ hybrid waveforms produced
with the BAM and 
Llama~\cite{Pollney:2009ut,Pollney:2009yz} codes are indistinguishable for an SNR 
below roughly 20. 

The mismatch error between two waveforms can be easily related to 
$||\delta h||$~\cite{McWilliams:2010eq}. 
If $\rho = \|h \|$ is the optimal SNR, then $||\delta h|| / \rho^2 \approx 2
{\cal M}$. This means that if two
waveforms meet the detection criteria of ${\cal M} < 0.03$, then they will be indistinguishable 
for $\rho < 4$, which is too weak a signal to be detected. If we want the waveforms to be
indistinguishable at an SNR of $\rho > 8$, then the mismatch must be below about 0.8\%. 
We see, then, that the accuracy requirements for two waveforms to be indistinguishable
are in general far more stringent than those for detection. We will return to this point
in Sec.~\ref{sec:PE}. 

All of the results in this work will be with respect to the Advanced LIGO 
detector~\cite{Abadie:2010cfa,AdLIGO}, and in
general we will use a low-frequency cut-off of 20\,Hz.

\section{Numerical waveforms}
\label{sec:waveforms}

We consider two families of black-hole binaries. %
The first is equal-mass binaries in which
the spin of each black hole is the same, $S_1/M_1^2 = S_2/M_2^2 = \chi$, 
and the spins are parallel or anti-parallel to the orbital angular momentum. In these 
configurations the spins do not precess, making this a simple subfamily of the 
black-hole-binary parameter space. The spins used were 
$\chi = \{0, \pm 0.25, \pm 0.50, \pm 0.75, \pm 0.85\}$, although we will only provide
length requirements up to $\chi = \pm 0.5$ for reasons that will be explained in 
Sec.~\ref{sec:lengthSpin}.
The second family is nonspinning binaries with mass ratios 
$q = M_2/M_1 = \{1,2,3,4\}$, where we labelled the individual masses such that
$M_2 \geq M_1$. 

The simulations were produced using the BAM 
code~\cite{Brugmann:2008zz,Husa:2007hp}. They cover 
6-10 orbits before merger.
The phase error of the waveforms is at most 0.15\,rad during
the inspiral phase (up to $M\omega = 0.1$), and on the order of 1\,rad during merger 
and ringdown. The amplitude accuracy is within 1\% during the inspiral, and within
5\% during the merger and ringdown. Of more relevance to GW detection
is the mismatch error of the waveforms, which is below $10^{-4}$ for all 
cases. Note that this is well within the detection requirement of 0.03, 
and indeed these waveforms are well within the accuracy 
requirements for both detection and parameter estimation with current
and planned ground-based detectors~\cite{Hannam:2009hh}.
These waveforms were all presented in detail 
in~\cite{Hannam:2007ik,Hannam:2007wf,Hannam:2010ec}.

\section{Hybrid inspiral-merger-ringdown waveforms}
\label{sec:hybrids}

Complete inspiral-merger-ringdown waveforms that include arbitrarily large
numbers of GW cycles before merger can be constructed by connecting PN and
NR results~\cite{Pan:2007nw,Ajith:2007qp,Ajith:2007kx,Boyle:2009dg,Santamaria:2010yb}. 
We model the inspiral regime using the time-domain approximants TaylorT1
and TaylorT4~\cite{Hannam:2007wf,Santamaria:2010yb}. For nonspinning binaries the phase 
evolution is described to 3.5PN order, and the amplitude is given to 3PN order~\cite{Blanchet:2008je}.
For spinning binaries, spin effects in the phase evolution are included only up to 2.5PN 
order~\cite{Hannam:2007wf}, and in the case of TaylorT4 we adopt two approaches
to truncating the expansion in a consistent 
way~\cite{Santamaria:2010yb,Hannam:2010ec}; the amplitude includes spin
contributions up to 2PN order~\cite{Arun:2008kb}.

The PN results are expansions in the frequency, $x = (M\Omega)^{2/3}$, where $\Omega$
is the orbital frequency of the binary motion. When decomposing the GW signal in spherical
harmonics, $\Omega$ is related to the frequency $\omega$ of the 
$(\ell=2,m=\pm2)$ modes by $|\omega| = 2 \Omega$. As such, the 
PN waveforms are most accurate for small $x$, i.e., many orbits before merger. As
the binary approaches merger, the errors in the PN approximation grow. 
The convergence properties of the 
PN expansion are not fully understood, and it is not possible to provide clear error estimates. 
One way to gain some insight into the PN errors is to compare 
results from different PN approximants and at different PN orders, and we will return to this 
idea later. 

The only way to 
definitively quantify the PN errors is to compare with fully general relativistic results. 
This can be done over the (relatively small) number of GW cycles for which we have
both PN and NR waveforms. PN-NR comparisons of phase and amplitude have been
performed over $\sim 10$ orbits prior to 
merger~\cite{Buonanno:2006ui,Baker:2006ha,Hannam:2007ik,Boyle:2007ft,Gopakumar:2007vh}.
Comparisons for the cases we
consider here are given in~\cite{Hannam:2010ec}. There we found that the TaylorT4
approximant has an accumulated phase error of no more than 0.2\,rad for nonspinning
cases, but the phase error grows to up to 2\,rad for equal-mass cases with large spin, 
in particular the $\chi < 0$ cases. The behaviour of the TaylorT1 phase appears to be most 
consistent across the parameter space that we consider, with an accumulated phase
error of about 1\,rad in all cases, although in most cases a version of TaylorT4 is more
accurate. The PN amplitude (at the highest-known PN order)
has an error as high as 4\% in the $\chi = 0.85$ case, and drops to around 2\% in 
the $\chi = -0.85$ case, while for nonspinning configurations it is around 3\%. (Note
however that the uncertainty in the NR amplitude for all of these cases is
1\%.) The existence of this region of good agreement between PN and NR results 
is what allows us to combine the two into hybrid waveforms. 

We construct time-domain hybrid waveforms by first decomposing the PN and NR waveforms
into their amplitude $[A_{\rm PN}(t)$ and $A_{\rm NR}(t)]$ and phase 
$[\phi_{\rm PN}(t)$ and $\phi_{\rm NR}(t)]$.
From the phase we can in turn calculate the frequency, $\omega(t) = \partial_t \phi(t)$. 
We then choose a matching frequency, $\omega_m$, and determine the time
when both $\phi_{\rm PN}$ and $\phi_{\rm NR}$ reach that frequency, and connect the two phase
functions at that time.  We independently determine the time when the PN and NR amplitudes
agree, and connect the amplitudes at that time. In this procedure we do not require that the 
transition between the PN and NR phases and amplitudes is any more than 
continuous. This procedure is
performed on $\Psi_4$. In mismatch calculations we 
first calculate $\tilde{\Psi}_4$ by making a fast Fourier transform (FFT) of
$\Psi_4$, and then produce $\tilde{h}$
by dividing by $-\omega^2$.

\subsection{Errors due to the hybridization process}
\label{sec:hyberrors}

There are currently several procedures to produce hybrid waveforms that have been
used in the literature, based either on a matching in the time or Fourier 
domains~\cite{Pan:2007nw,Ajith:2007qp,Ajith:2007kx,Boyle:2009dg,Santamaria:2010yb}.
Each procedure will itself introduce artifacts into the waveform, and we may be concerned
that these artifacts will be a large source of error. Here we will compare hybrids 
produced using three different methods, and show that in fact the mismatch error 
introduced by the hybridization procedure is negligible. 

We consider three equal-mass nonspinning hybrids. The first is produced by the method 
that we have just described. The second is produced using the method described 
in~\cite{Ajith:2007qp,Ajith:2007kx}, but applied to $\Psi_4$ instead of the wave strain
$h$, because we want to assess the errors due to the hybridization process, without
any contamination by errors that may be introduced by a time-domain integration of 
$\Psi_4$ to $h$. Our specific method is to choose
a matching frequency, $\omega_m$,
and to then locate the time in both the PN and NR waveforms when that frequency is
reached. We then combine the two waveforms over a $200M$-long window, aligning
the waveforms such that the quantity 
\begin{equation}
\Delta \Psi = \int_{t_0}^{t_1} \left( \Psi_{4, \rm NR}(t) - a \, e^{i \delta \phi} \,
\Psi_{4, \rm PN}(t+ \delta t) \right)^2 dt
\end{equation} is minimized, where $t_0$ and $t_1$ are respectively $100M$ before and 
after the time $t_m$ at which each waveform reaches $\omega_m$, 
$a$ is a scale factor, $\delta t$ and $\delta
\phi$ are time and phase offsets, and the waveforms are initially aligned so that 
$t_m$ is the same for both. The hybrid is constructed by making a linear
transition between
$\Psi_{4, \rm PN}$ and $\Psi_{4, \rm NR}$ over the matching window. 
For both hybrids, we choose a matching frequency of $M\omega_m = 0.07$. 

The third hybrid is constructed in the frequency domain, 
using a variant of the method described in~\cite{Santamaria:2010yb}. We produce an FFT of the 
time-domain TaylorT4 approximant, and an FFT of the numerical $\Psi_4$ data. 
The phase of the frequency-domain PN and NR signals is then 
matched in the window $M\omega \in [0.0566, 0.113]$ and the continuous
transition is carried out at the matching frequency $M\omega_m = 0.079$. 

\begin{figure}[t]
\centering
\includegraphics[width=90mm]{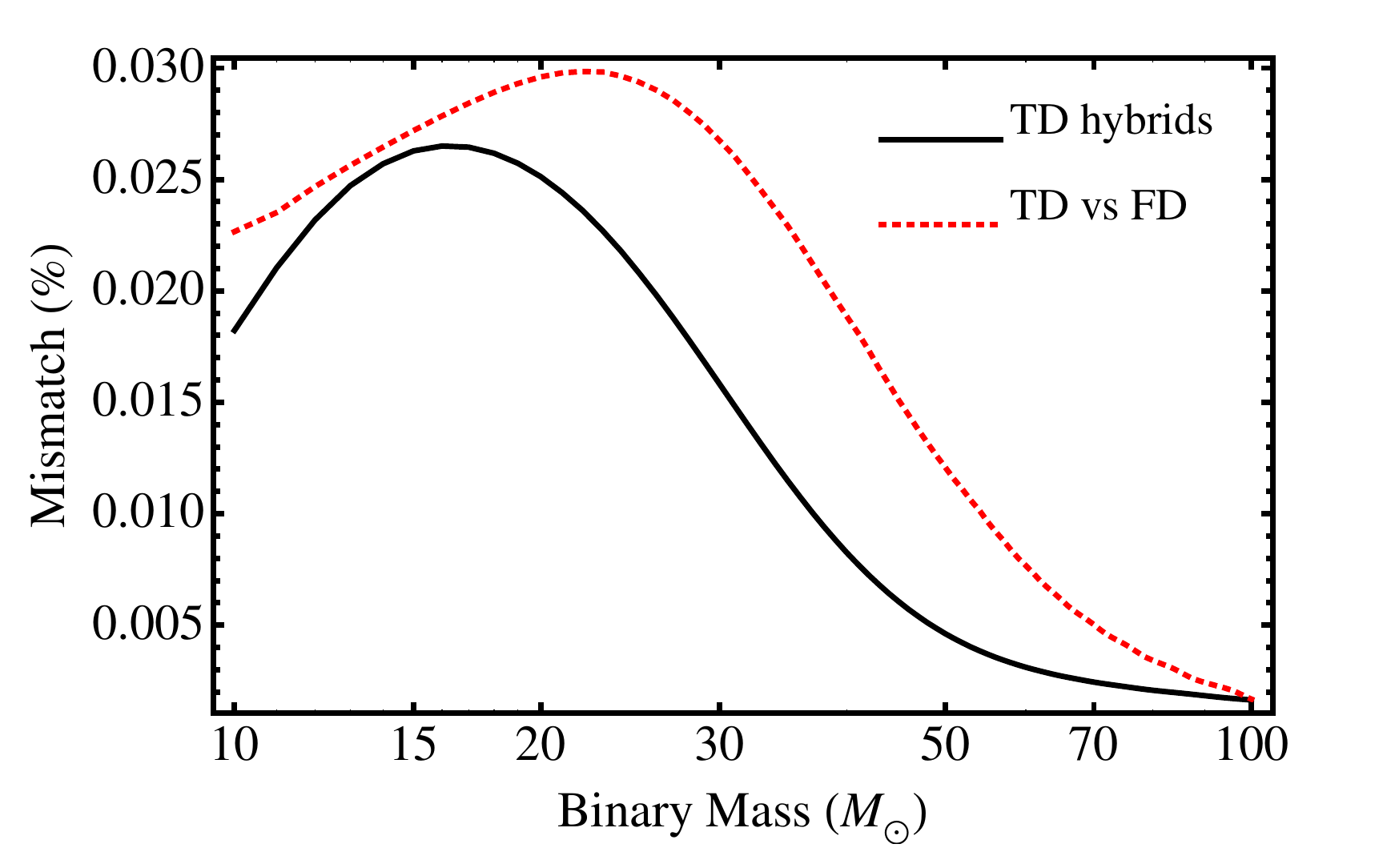}
\caption{
Mismatch between equal-mass nonspinning T4+NR hybrids constructed using two 
different time-domain methods (solid line), and between hybrids constructed in the time 
domain (TD) and the frequency domain (FD). The time-domain hybrids are produced
with matching frequency $M\omega_m = 0.07$, and the frequency-domain hybrid
is matched at $M\omega = 0.079$. 
}
\label{fig:hybcomparison}
\end{figure}

The non-mass-optimized mismatch
is shown in Fig.~\ref{fig:hybcomparison}. 
We see that the maximum mismatch is about 0.025\% between the two time-domain 
hybrids, and 0.03\% between the time-domain and frequency-domain hybrids.
Clearly the error due purely to the hybridization
procedure is, like the mismatch error of the numerical waveforms, negligible. 

If we consider the indistinguishability criteria, $||\delta h||<1$, then these hybrids
would be indistinguishable for SNRs of $\rho < 40$. 

We note, however, that the difference between hybrids constructed with 
ostensibly the same numerical waveforms and PN approximants may have 
larger differences than those shown here. For example, if we compare either
of the hybrids we have just described, with hybrids constructed using the 
integrated wave strain, as in~\cite{Ajith:2007qp,Ajith:2007kx}, then the mismatch
can be as high as 0.8\%. This is due not to the 
hybridization process, but to artifacts introduced in the time-domain integration
of $\Psi_4$ to $h$.


\section{Comparison between PN and NR during the late inspiral}
\label{sec:T1hybrids}

Direct comparisons between the phase and amplitude of PN approximants and
NR waveforms have been made 
in~\cite{Buonanno:2006ui,Baker:2006ha,Hannam:2007ik,Boyle:2007ft,Gopakumar:2007vh,Hinder:2008kv,Campanelli:2008nk}, 
and we refer the reader to~\cite{Hannam:2010ec} for PN-NR comparisons of the binary 
configurations studied here. In this section we reframe those comparisons in the context of
hybrid waveforms and mismatches. 

For several choices of binary configuration, we 
construct a fiducial reference hybrid waveform by matching a 3.5PN TaylorT1 waveform to 
the longest available NR simulation. We then construct 
a set of ``candidate'' hybrid waveforms by sliding the matching region over the NR 
waveform (such that progressively shorter NR segments are employed in the hybrid 
construction). How well one candidate hybrid waveform can be distinguished from 
the reference waveform is given by the mismatch between the ``reference'' 
and the ``candidate''. 

We study three cases, employing three different NR simulations: an equal-mass, 
non-spinning simulation performed using the \texttt{SpEC} 
code~\cite{Scheel:2008rj,SpEC:wfs}, 
and two of the BAM simulations presented in~\cite{Hannam:2010ec},
equal-mass simulations with spins aligned and anti-aligned to the orbital angular 
momentum  ($\chi = \pm0.75$).
Note that in this section, we choose a low-frequency cut-off of 10\,Hz, the mismatches are \emph{not} maximized over the binary mass, 
and the hybrids are constructed using the procedure described in~\cite{Ajith:2007qp,Ajith:2007kx}.

\begin{figure}[tb]
\centering
\includegraphics[width=3.5in]{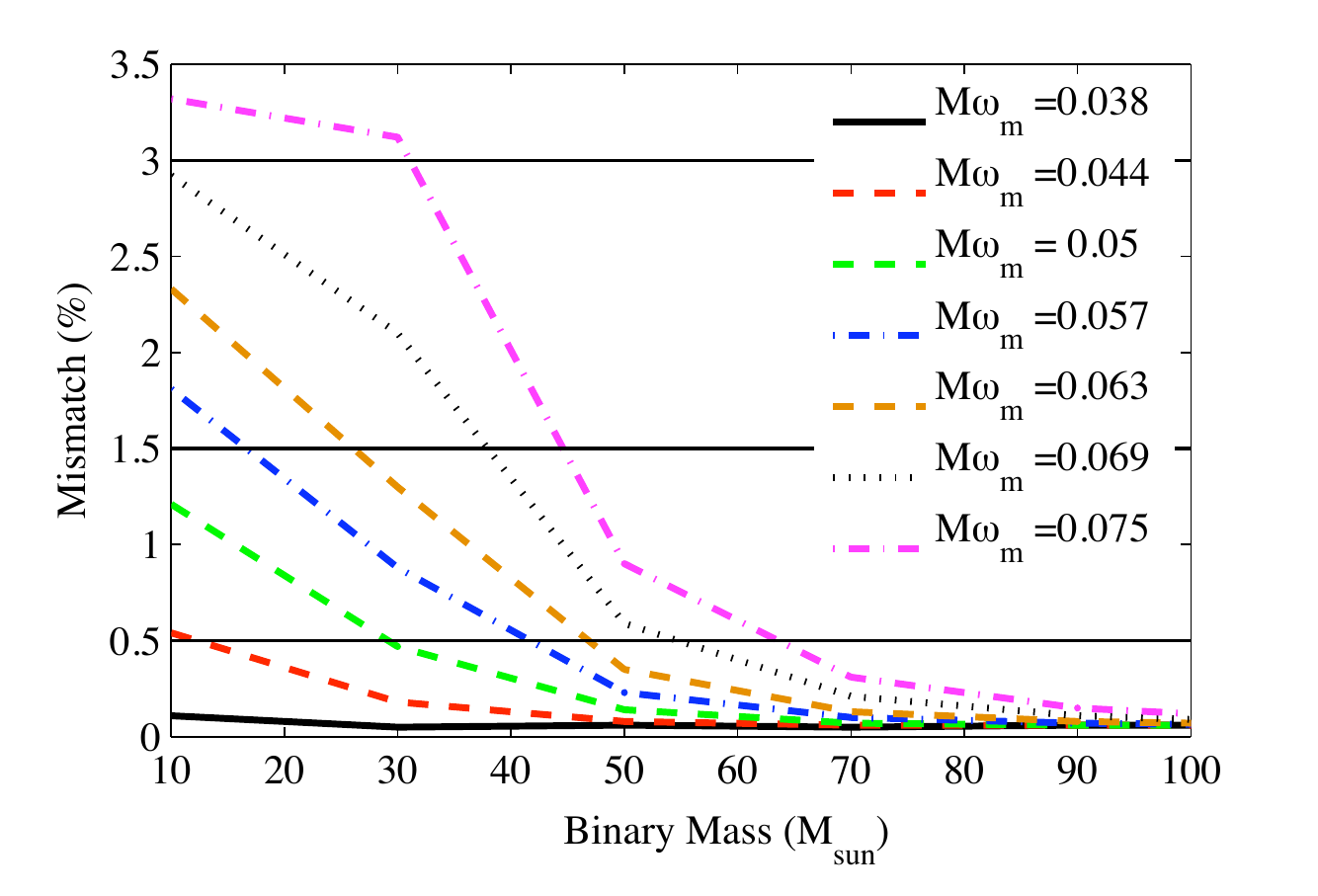}
\caption{Mismatch between the \emph{reference} hybrid waveform and the hybrid
waveforms constructed using different matching frequencies. 
The hybrid waveforms are constructed by matching TaylorT1 PN waveforms
with the Caltech-Cornell equal-mass, non-spinning NR simulation.}
\label{fig:CaltechWaveform}
\end{figure}

Figure~\ref{fig:CaltechWaveform} shows the mismatch between the reference hybrid 
waveform and the hybrid waveforms constructed using different matching regions 
in the equal-mass, non-spinning case.

The matching frequency of the reference waveform is $M\omega_{\rm m,r} = 0.031 - 0.038$. 
We will denote the matching frequency of the candidate hybrid by $\omega_{\rm m}$.

When $\omega_{\rm m,r} = \omega_{\rm m}$, then the two waveforms are identical, 
and the mismatch is zero. As the matching frequency of the candidate waveform 
$\omega_{\rm m}$ is increased, the mismatches grow. This gives us a picture of the 
mismatch due to the disagreement of TaylorT1 and full numerical results over the 
frequency range  $M\omega \in [\omega_{\rm m,r},\omega_{\rm m}]$. For the maximum
matching frequency we choose, $M\omega_{\rm m} = 0.075$, the maximum mismatch
is 3.5\%.

\begin{figure}[tb]
\centering
\includegraphics[width=3.5in]{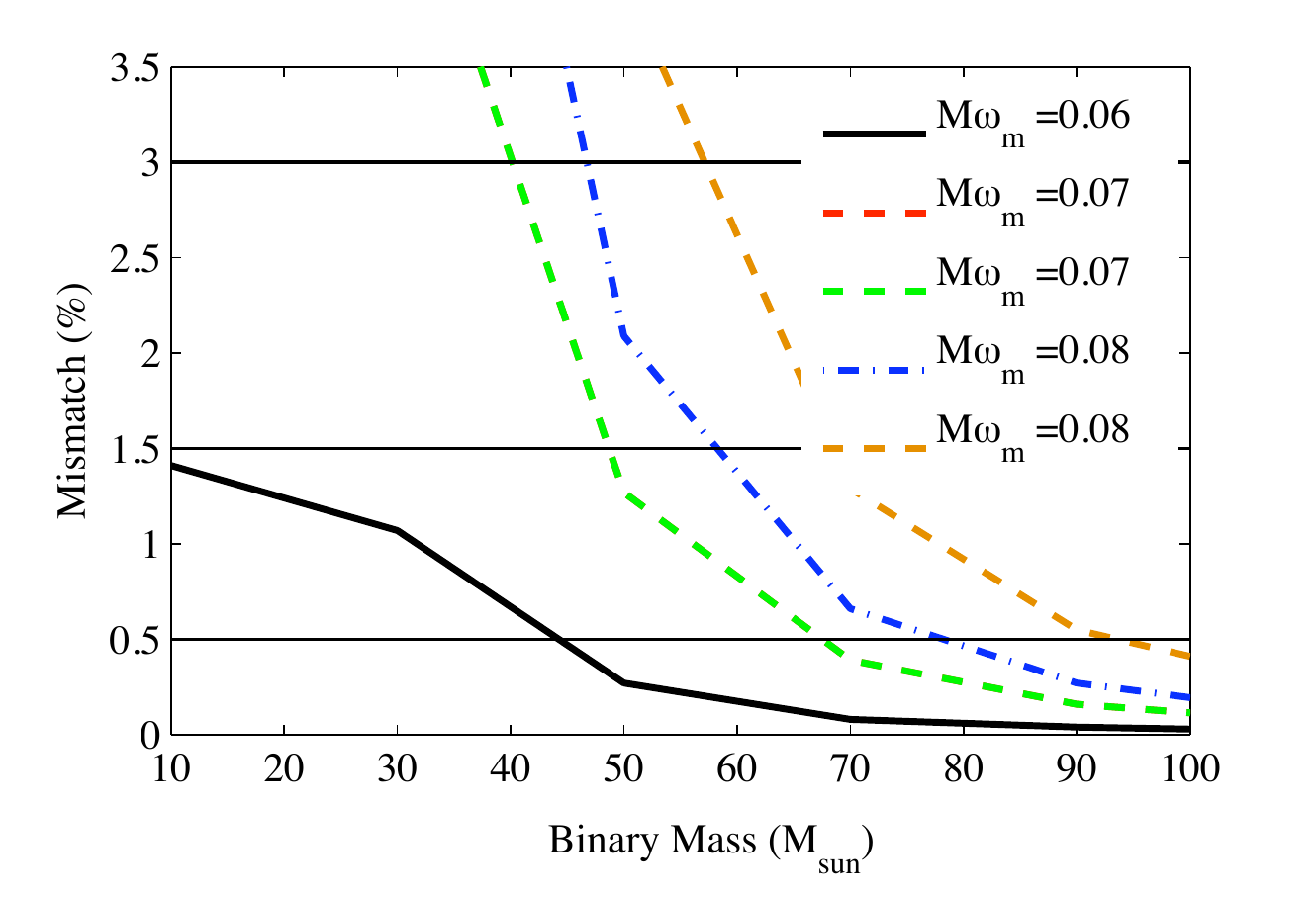}
\caption{Same as in Fig.~\ref{fig:CaltechWaveform} except that the hybrid
waveforms are constructed by matching TaylorT1 waveforms with BAM simulation
of equal-mass binary with spins $\chi = 0.75$.}
\label{fig:BAMWaveformSpin0.75}
\end{figure}

\begin{figure}[tb]
\centering
\includegraphics[width=3.5in]{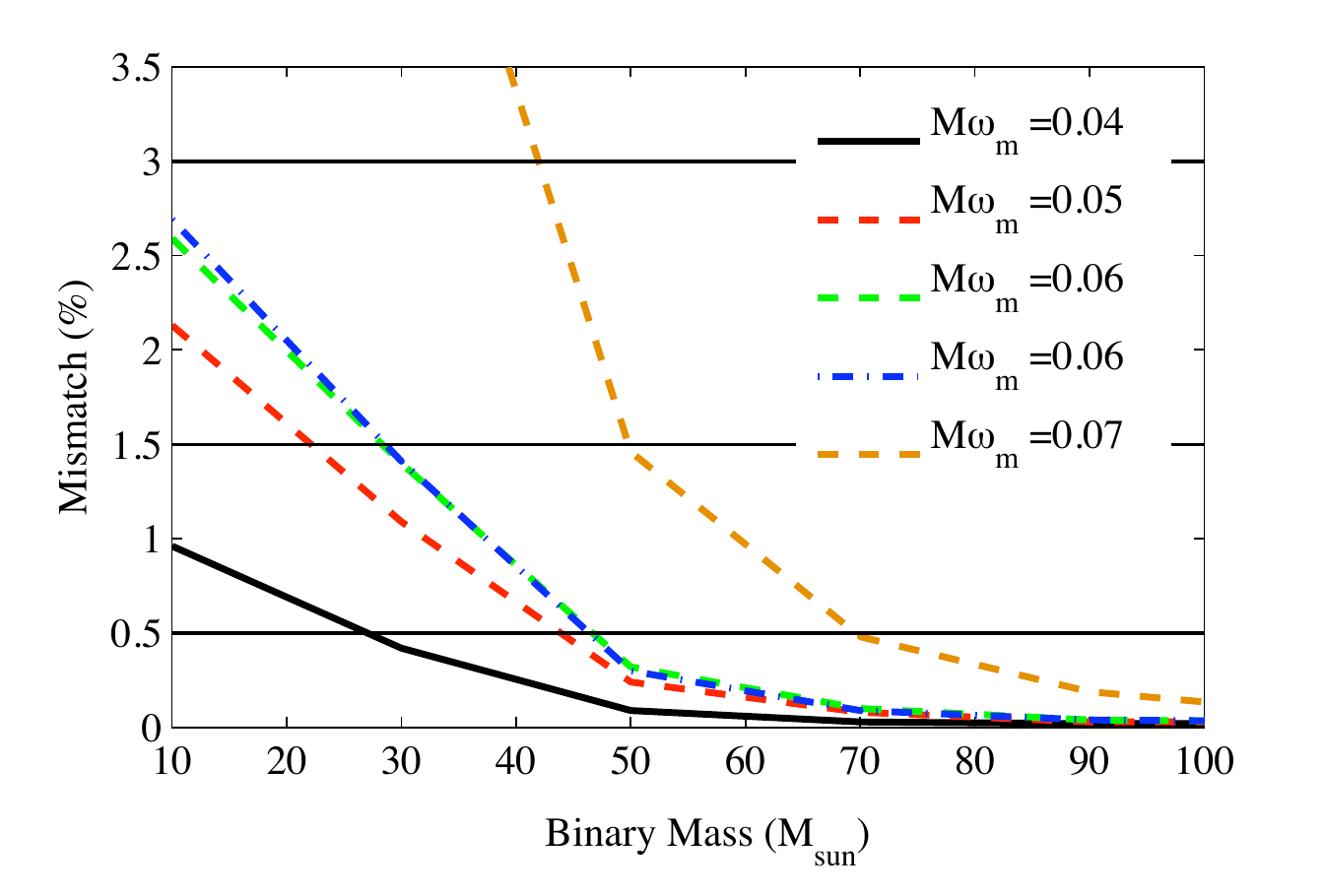}
\caption{Same as in Fig.~\ref{fig:CaltechWaveform} except that the hybrid
waveforms are constructed by matching TaylorT1 waveforms with BAM simulation
of equal-mass binary with spins $\chi = -0.75$.}
\label{fig:BAMWaveformSpin-0.75}
\end{figure}

Figures~\ref{fig:BAMWaveformSpin0.75} and~\ref{fig:BAMWaveformSpin-0.75} show the
mismatches between the reference and candidate hybrid waveforms for the case of 
equal-mass binaries with aligned and anti-aligned spins $\chi = \pm 0.75$, respectively. 
The $\chi = -0.75$ reference waveform was constructed with a matching frequency of 
$M\omega_{\rm m,r} \simeq 0.038 - 0.044$, 
and the $\chi = 0.75$ waveform is matched at $M\omega_{\rm m,r} \simeq 0.057 -
0.063$.

We see from these results that the mismatch rises above 3\% at lower matching frequencies 
than in the non-spinning example. This is a reflection of the fact that the TaylorT1 approximant
performs worse in the spinning cases. 

These first results give us an indication of the mismatch error associated with the 
disagreement between the TaylorT1 approximant and NR results over the 
frequency range for which NR results exist. They show that the accumulation of
mismatch error over that frequency range due to PN errors is orders of magnitude
larger than the mismatch error due to errors in the numerical waveforms or to the 
hybridization process.
These mismatches could drop significantly if we were to minimize the mismatch 
with respect to binary mass, but that would not alter the qualitative observation that
PN errors dominate the error in our hybrid waveforms.


\section{Waveform length requirements: general procedure}
\label{sec:length}

In producing hybrid waveforms, we have seen that the dominant error source
is the error in the PN approximation. 
We expect the PN waveforms to be 
increasingly accurate as we move to lower frequencies (i.e., earlier in the inspiral), 
and so the lower the frequency $\omega_{\rm m}$ at which we can match 
(i.e., the more NR cycles we
can use) the better. Our goal now is to determine the highest acceptable
matching frequency for a variety of accuracy criteria for the full waveforms. 

A previous study showed that pure PN approximants of nonspinning binaries 
(i.e., without any stitching to merger and ringdown signals), are adequate for detection 
purposes up to about 12\,$M_\odot$~\cite{Buonanno:2009zt}. 
Therefore our PN+NR hybrids have to be accurate only for total masses $M \apprge
10\,M_\odot$. Above these
masses the mismatch between PN inspiral waveforms and full
 inspiral-merger-ringdown waveforms deteriorates 
 quickly~\cite{Pan:2007nw,Buonanno:2009zt,Bose:2009}.

The main difficulty in assessing the physical accuracy of hybrids is that, since we cannot 
rigorously quantify error bars in the PN waveforms, any error estimates we make on our complete waveforms can be no
better than an educated guess. The best we can do is to make clear what 
assumptions have gone into our guess, and to what extent this guess can be considered
conservative or optimistic. We will now describe the procedure we have used, 
and our justifications for it. 

Our approach is the following. 
We start by defining a model waveform that we will regard for the purposes of 
this study as the ``true'' physical waveform. 

For nonspinning 
binaries we know that during the frequency range in which PN and NR results overlap, 
the TaylorT4 approximant captures the phase evolution with high accuracy.
We now {\it assume} that the TaylorT4 phase continues to be physically correct 
to much lower frequencies, and we construct a 
TaylorT4+NR hybrid and treat that as the true GW waveform from our system. 
In other words, we treat this hybrid as if it were a numerical waveform starting
at $M\omega = 0.006$, which corresponds to 20\,Hz for a 10\,$M_\odot$ 
binary. Such a waveform covers $\sim$650 cycles before merger. We have no 
expectation of ever producing such a waveform in a numerical simulation, 
but for the purposes of this exercise we will treat our hybrid as if it were the result of 
just such a  simulation, and refer to it as NR$_{\rm L}$. 

Our goal is to assess the accuracy of PN+NR hybrids
constructed using a \emph{reasonable} choice of PN approximant. Based on its 
phase accuracy near merger, we choose the TaylorT1 approximant. 
We construct hybrids of TaylorT1 with NR$_{\rm L}$, and due to the great length
of NR$_{\rm L}$ we have the freedom of matching at arbitrarily low frequencies.

We then calculate the mismatch between T1+NR$_{\rm L}$ and the true 
NR$_{\rm L}$ waveform, and use this to estimate the ability of T1+NR$_{\rm L}$ 
to detect the true signal. This process is repeated for a range of matching 
frequencies, which allows us to determine the highest matching frequency at 
which the T1+NR$_{\rm L}$ hybrid is sufficiently accurate for detection of 
binaries above 10\,$M_\odot$. 

The one key assumption in this procedure is that the TaylorT4 phase
can be trusted to much lower frequencies than those where it has currently been 
compared with full numerical results. The first and most detailed study of TaylorT4 was 
presented in~\cite{Boyle:2007ft}, which included frequencies
down to $M\omega \approx 0.035$, so we know that TaylorT4 (at least in the nonspinning
case) is accurate down to that frequency. But below that frequency we have no information 
about its accuracy with respect to waveforms from full general relativity. 
We expect that the {\it difference} between the TaylorT4 and TaylorT1 phases provides a reasonable estimate of the phase error in a typical approximant, but there is no way to prove this.
Ultimately one is reduced to a statement of faith in the accuracy of PN methods. 

Let us illustrate this point with two extreme views. 

Instead of comparing TaylorT1 and TaylorT4 hybrids, we could compare hybrids produced
using different effective-one-body (EOB) calculations. It has been claimed that EOB 
methods, particularly after calibration to numerical results, provide an extremely accurate
model of the full GW 
signal~\cite{Buonanno:1998gg,Buonanno00a,Damour:2000we,Damour:2001tu,Buonanno:2005xu,Buonanno:2007pf,Damour:2007yf,Damour:2007vq,Damour:2008te,Baker:2008mj,Mroue:2008fu,Damour:2009kr,Buonanno:2009qa,Pan:2009wj}. 
As with standard PN approximants, EOB results can
only be tested in the regime where NR results exist. However, all PN and EOB 
results make use of expressions for the GW flux and energy loss of the binary, 
and it has been shown in~\cite{Boyle:2008ge} that EOB estimates of these
quantities agree with NR results far better than {\it any} standard PN estimates.
We might then expect, on this evidence alone, that EOB results are far more 
accurate than any standard PN approximant, and the appropriate comparison
would be between hybrids produced using variants of the EOB method.

This view suggests that the comparison we are proposing --- between T1 and T4
hybrids --- is overly pessimistic of the potential physical
fidelity of hybrid waveforms. 

An alternative view is that  EOB waveforms (whether the final waveform phase, or the
flux and energy-derivative ingredients) have nonetheless only been compared with full GR
results close to merger, and their accuracy at lower frequencies is unknown. One
could argue that we should be much {\it more} conservative, and instead of comparing
3.5PN TaylorT1 and TaylorT4 hybrids, we should compare hybrids constructed
using different PN orders. Only by comparing, for example, 2.5PN and 3.5PN results,
can we hope to estimate the error between 3.5PN results and those from full GR. As
we will see in Sec.~\ref{sec:lengthUM}, this provides an extremely pessimistic view 
of the accuracy of 3.5PN waveforms. 

In the end, then, we consider our proposed T1-T4 comparison to be a reasonable 
compromise between these two extreme views. But we hope this long preamble serves
as a caveat that all such comparisons are by no means fully conclusive.

\section{Nonspinning binaries}
\label{sec:lengthUM}

We will now investigate the length requirements for simulations of 
nonspinning binaries. We start with the equal-mass nonspinning case. 

\begin{figure}[t]
\centering
\includegraphics[width=90mm]{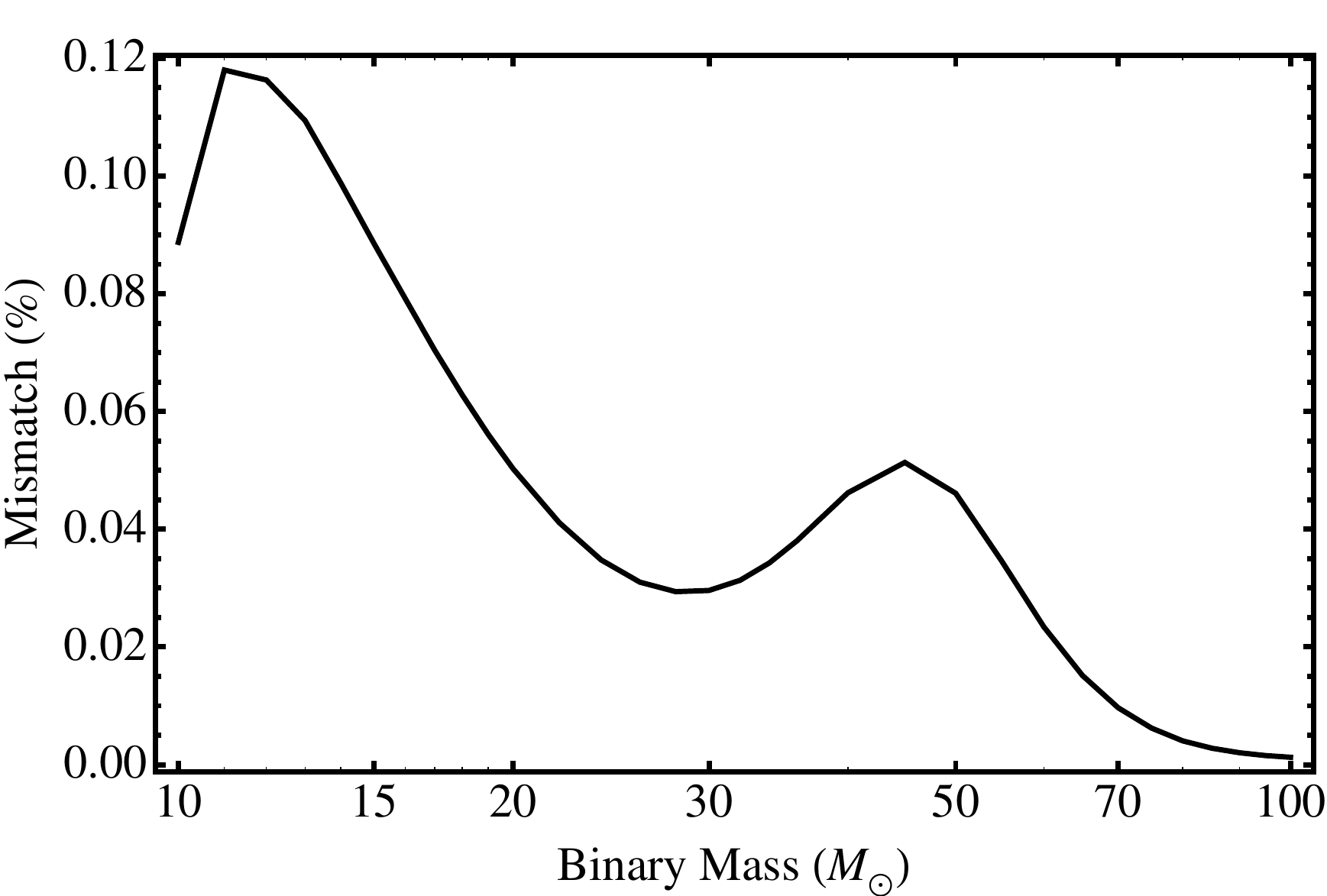}
\caption{
Mismatch between T4+NR hybrids constructed with matching frequencies
$M\omega_m = 0.06$ and $M\omega_m = 0.09$. We see that the mismatch
between T4 hybrids constructed at different matching frequencies is below 0.12\%
over the entire mass range that we consider.
}
\label{fig:T4mismatches9060}
\end{figure}

We first construct a hybrid between 3.5PN TaylorT4 and our numerical 
waveform. We make the transition between PN and NR at a matching frequency
$M\omega_m = 0.09$. Fig.~\ref{fig:T4mismatches9060} shows the mismatch between 
this waveform and a
hybrid produced with a different matching frequency, 
and suggests that for our purposes the resulting hybrid does not 
depend strongly on the matching frequency $\omega_m$, since we are 
not concerned by mismatches that are below 0.5\%. Note that these results
are a further reflection of the good agreement of TaylorT4 with the numerical phase 
at these frequencies; we see from Fig.~\ref{fig:CaltechWaveform} that the same would
not be true for T1+NR hybrids.

We treat this waveform as if it were an extremely long 
numerical waveform, and denote it NR$_{\rm L}$. This is our 
target GW signal. It is
around $4\times10^5M$ in duration, making it two orders of magnitude longer
than the longest current equal-mass nonspinning 
waveform~\cite{Boyle:2007ft}. TaylorT1+NR$_{\rm L}$ hybrids
are then constructed with a range of matching frequencies, and compared
with NR$_{\rm L}$, as described in the previous section.

\begin{figure}[t]
\centering
\includegraphics[width=90mm]{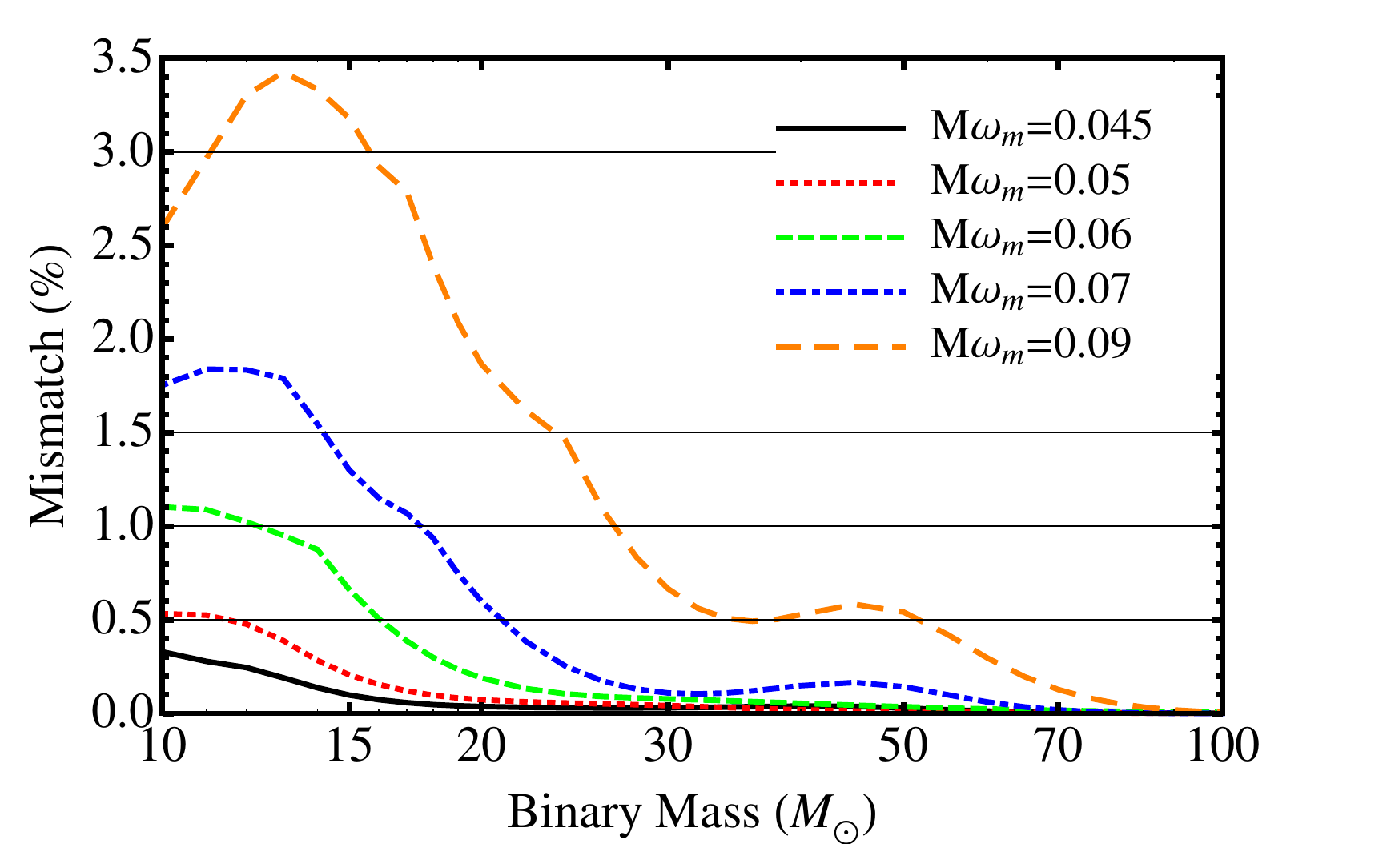}
\caption{
Mismatches between T1+NR$_{\rm L}$ and NR$_{\rm L}$ hybrids, for
matching frequencies $M\omega_m = \{0.045,0.05,0.06,0.07,0.09\}$. 
}
\label{fig:q1mismatches}
\end{figure}

Fig.~\ref{fig:q1mismatches} shows the mismatch between 
NR$_{\rm L}$ and T1+NR$_{\rm L}$ hybrids
for $M\omega_m = 0.045, 0.05, 0.06, 0.07, 0.09$. 
Above the highest matching frequency of $M \omega_m = 0.09$ both hybrids 
are identical for all the matching frequencies
we have chosen. This means that for masses above about 150\,$M_\odot$, the
two waveforms will be almost identical within the sensitivity range of the detector,
and their mismatch should approach zero. For this reason Fig.~\ref{fig:q1mismatches} 
considers masses only up to 100\,$M_\odot$, and indeed we see that the
mismatch has essentially dropped to zero by $100\,M_\odot$. 

At lower masses, the difference between the two PN approximants dominates,
and the mismatch rapidly grows. At yet lower masses, the PN approximants are being
sampled at lower frequencies, where their agreement is better, and we expect
the mismatch to fall. This behavior can be seen for the matching frequency 
$M\omega_m = 0.09$, where the peak mismatch is at about 13\,$M_\odot$. 
For lower matching frequencies, the peak mismatch is pushed to lower masses,
and for $M\omega_m \leq 0.05$ we do not see it in the figure. 
We expect that the location of the mismatch maximum is related to the mass at
which 
$M\omega_m$ is at the detector's most sensitive frequencies. The detector's peak 
sensitivity is at around 200Hz, and for example $M\omega = 0.09$ corresponds to that 
frequency for $M=15.4\,M_\odot$, which is consistent with Fig.~\ref{fig:q1mismatches}.

As described in Sec.~\ref{sec:length}, the mismatch is optimized with respect to 
the total mass of the binary. We cannot optimize over the mass ratio, because
we do not have access to numerical waveforms with $q$ close to unity. 
However, we can make a crude estimate of the magnitude of the effect
of mass-ratio optimization. If we consider a 10\,$M_\odot$ binary, then most
of the hybrid in the detector's sensitivity band is from the PN contribution. We
vary the mass ratio of a TaylorT1 waveform until its phase agreement with 
TaylorT4 is optimized, and then construct a TaylorT1+NR$_{\rm L}$ hybrid
using that mass ratio, and repeat the above analysis with $M\omega_m = 0.06$.
The optimal mass ratio is $q = 0.99985$, and the change in the mismatch 
is indistinguishable on the scale of Fig.~\ref{fig:q1mismatches}. This suggests
that the mass optimization by itself is sufficient to give an indication of the full mismatch
error of our hybrids. 

If we are willing to consider waveforms acceptable so long as the waveform-error
mismatch is below 3\%, then we see from Fig.~\ref{fig:q1mismatches} that the 
$M\omega_m = 0.09$ T1 hybrids are 
almost good enough, and hybrids produced with any lower matching frequency 
are acceptable. With this criterion, we need waveforms that contain only five GW
cycles before merger, or about two orbits! (Note that the number of GW cycles before
the peak amplitude is more than simply twice the number of orbits, because the 
approximate quadrupole relation $\omega = 2\Omega$ no longer holds.)

Being more realistic, and demanding that the mismatch is below 1.5\% (so that
the remainder of our allowed 3\% mismatch is taken up by the template spacing),
a matching frequency of $M\omega_m = 0.06$ is more than acceptable. This means
that we need about 11 GW cycles before merger, or about five orbits. If we
are especially stringent and require that the error mismatch be below 0.5\%, then
we must match below $M\omega_m = 0.05$, and this requires 15 cycles 
(seven orbits). 

For the equal-mass nonspinning case, then, we see that the NR waveform length
requirements for GW detection are not very high: five orbits are sufficient,
and seven orbits are plenty.

\begin{figure}[t]
\centering
\includegraphics[width=90mm]{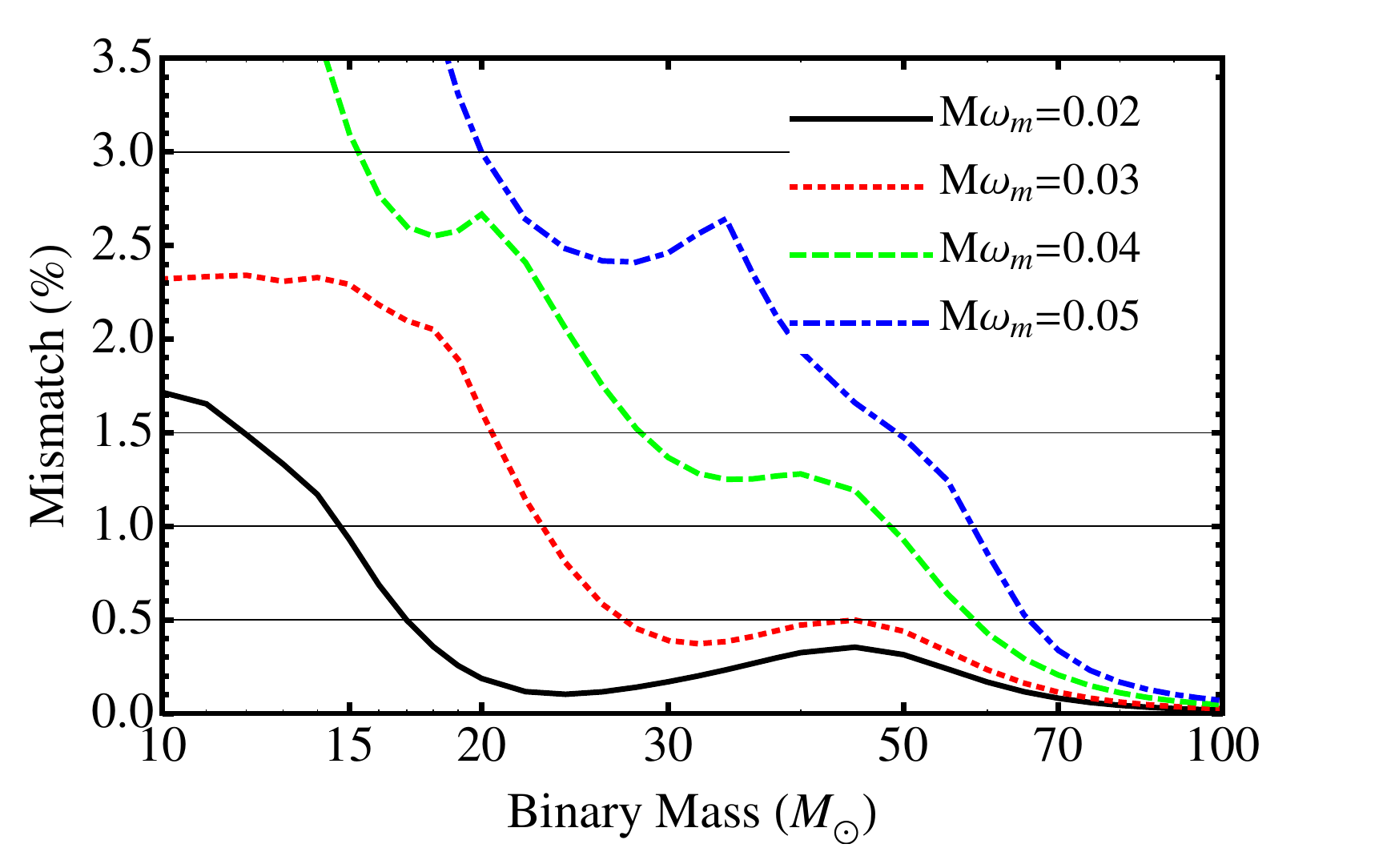}
\caption{
Mismatches between T1+NR$_{\rm L}$ and NR$_{\rm L}$ hybrids, where
the TaylorT1 approximant is evaluated at only 2.5PN order.
}
\label{fig:q1mismatches25PN}
\end{figure}

At this point we may ask what happens if we decide to be much more conservative in 
estimating the PN error. We can do this by repeating the exercise above, but instead
using TaylorT1 waveforms at only 2.5PN order. The results are shown in 
Fig.~\ref{fig:q1mismatches25PN}. We see that the variation in the mismatch is dramatic. 
Even if we choose a matching frequency of $M\omega_m = 0.04$ (which corresponds to
23 GW cycles before merger), the waveforms are not accurate enough for searches,
irrespective of template bank spacing, for masses below 15\,$M_\odot$. Only if
we match at $M\omega = 0.03$ (40 cycles before merger) are the waveforms usable,
and to get close to the more rigorous mismatch requirement of 1.5\% for all
masses, we need $M\omega = 0.02$, which corresponds to 80 GW cycles before
merger. If one is to adopt the view that this is a realistic estimate of the uncertainty 
in our 3.5PN approximants, then we must conclude that hybrids constructed from 
even the longest numerical waveforms currently published are only useful for 
searches down to 15-20 solar masses. It is the opinion of the authors, however, 
that this is a gross exaggeration of the error in 3.5PN approximants!

\begin{figure}[t]
\centering
\includegraphics[width=90mm]{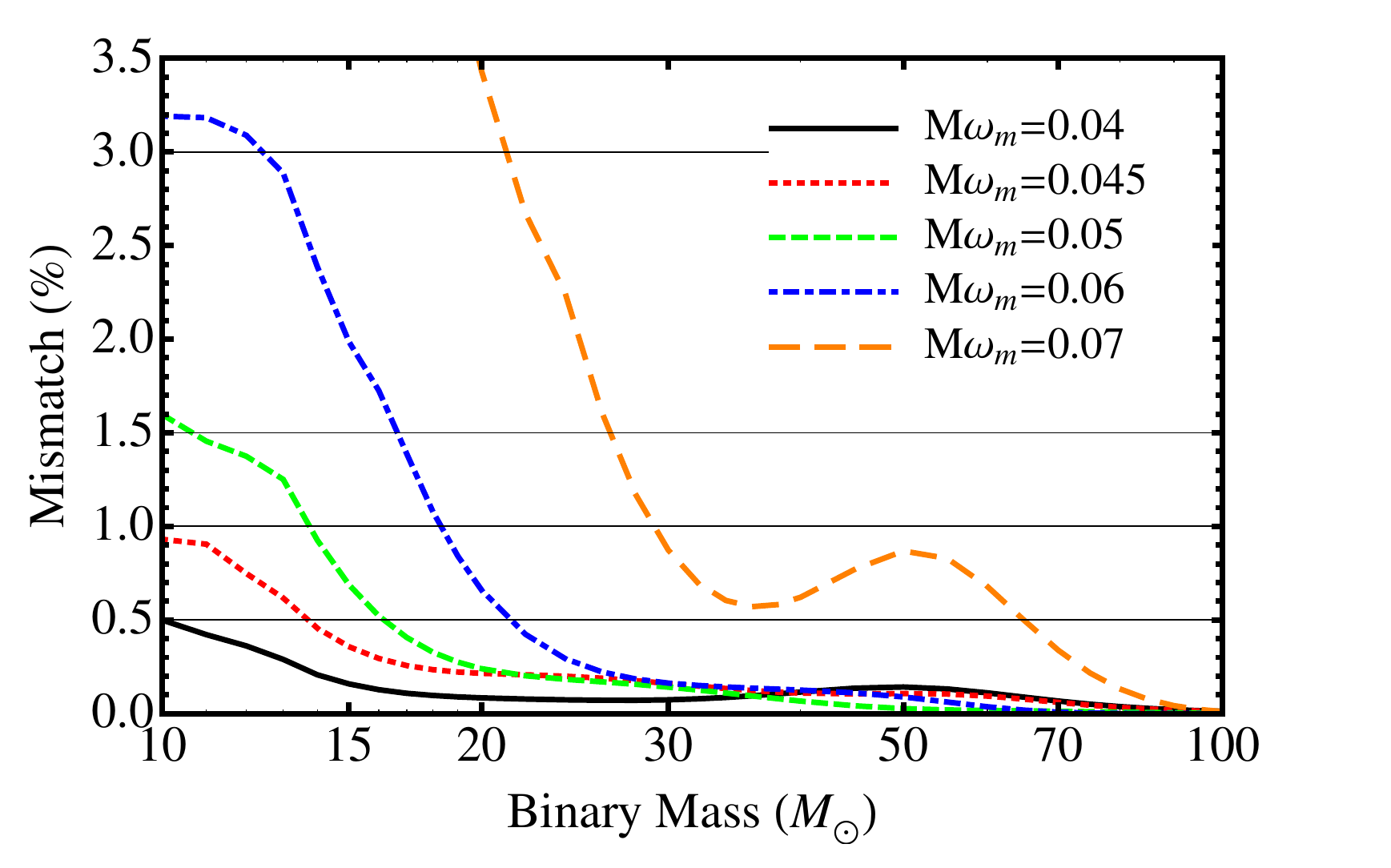}
\caption{
Mismatches between T1+NR$_{\rm L}$ and NR$_{\rm L}$ hybrids for $q=4$, with
matching frequencies $M\omega_m = \{0.04,0.045,0.05,0.06,0.07\}$.}
\label{fig:q4mismatches}
\end{figure}

We now consider higher mass ratios. The results in~\cite{Hannam:2010ec} suggest
that TaylorT4 is also a good model for the true waveform for the mass ratios that 
we consider, $q = \{1,2,3,4\}$. We find that the maximum acceptable matching frequency
drops as the mass ratio increases, so that we need longer numerical waveforms. 
Fig.~\ref{fig:q4mismatches} shows the results for $q=4$, which is the most extreme
case. The figure shows results with matching frequencies 
$M\omega_m = \{0.04,0.045,0.05,0.06,0.07\}$. Clearly matching frequencies of 
$M\omega = 0.07$ will not be sufficient even if we allow mismatches up to
3\%, but $M\omega_m = 0.06$ (15 cycles, or about 7 orbits before merger) is borderline.  
A matching frequency of $M\omega_m = 0.05$ (21 cycles, 10 orbits) ensures mismatches no 
higher than 1.5\%, and mismatches below 0.5\% require a matching frequency of
$M\omega = 0.04$ (33 cycles, or about 15 orbits). 

We see, then, that for higher mass ratios it is not sufficient to produce waveforms of 
the same modest lengths as in the equal-mass case. This is unfortunate, because
higher-mass-ratio simulations are far more computationally expensive. This result
also highlights a drawback of direct comparisons between NR and PN results
over a small number of cycles before merger: in such comparisons, the performance
of TaylorT1 and TaylorT4 is roughly the same for $q=1$ and $q=4$, and we
might therefore conclude that the length requirements for NR waveforms would 
be the same if we want to produce sufficiently accurate TaylorT1+NR hybrids
in either case. But these results show otherwise; we need more cycles for
higher mass ratios. 

The $q=4$ numerical waveform used for this study covers 17 GW cycles before 
merger, and so would in principle be usable for searches down to 10\,$M_\odot$, although 
a slightly longer waveform (for example including 10 clean inspiral cycles) would be
preferable. It is clear, however, that (1) producing acceptable $q=4$ waveforms is 
certainly feasible with current codes, but (2) it is non-trivial to estimate the length
requirements for yet higher mass ratios.

\section{Equal-mass spinning binaries}
\label{sec:lengthSpin}

We now tackle the cases with spin. The TaylorT4 approximant no longer accurately 
tracks the true phase in the last cycles before merger, and it would be difficult to justify its use 
in constructing an NR$_{\rm L}$ hybrid as in the previous section. (See Figs.~7 and 8 
in~\cite{Hannam:2010ec}.)

Without a promising approximant to use to produce a stand-in for a true waveform, 
our next-best option is to construct one. The nonspinning terms of TaylorT4 are known
up to 3.5PN order, but the spinning terms are known only up to 2.5PN. Perhaps if 
we knew the 3PN and 3.5PN contributions, the spinning TaylorT4 would be just 
as good as the nonspinning version? Whether this is true or not, we can certainly 
introduce 3PN and 3.5PN terms that improve its agreement with numerical results. 
Our procedure is to adjust the coefficients of such terms so as to minimize the 
square-integral phase difference between the PN and NR waveforms, as defined
by~\cite{Hannam:2010ec} \begin{equation}
\overline{\Delta \phi}(t_{N}) =  \frac{1}{\sqrt{-t_{N}}} \left[ \int_{t_{N}}^{0} \Big( \phi_{\rm NR}(t) - \phi_{\rm PN}(t) \Big)^2 dt \right]^{1/2}, \label{eqn:sqrdphi}
\end{equation} where $N$ is the number of cycles included in the comparison, and 
we choose $N=10$. We find in practice that the phase evolution 
is almost identical using an approximant with either an optimized 3PN coefficient only, 
or both 3PN and 3.5PN coefficients. This suggests that our fitting procedure over 10 GW 
cycles is not very sensitive to higher order coefficients.

\begin{table}
\caption{\label{tab:PNcoeffs}
Empirically calculated 3PN and 3.5PN coefficients that produce a TaylorT4
approximant that agrees well with the NR phase over the last ten cycles
up to $M\omega = 0.1$.  
}
\begin{tabular}{|| c | c | c ||}
\hline
$\chi$ & 3PN & 3.5PN \\
\hline
-0.85      & 989.4   & 16.26     \\
-0.75      & 775.3   & 21.11     \\   
-0.50      & 259.2   & 23.99     \\
-0.25      & 39.73   & -1.22      \\
0             & 0.868   & -0.111    \\
0.25       & -128.9  & 0.320 \\
0.50       & -204.6  & 1.070 \\
0.75       & -298.5 & 38.97 \\
0.85       & -268.6 & -1.009   \\
\hline
\end{tabular}
\end{table}

However, the minimization process is sensitive to the initial guess
for the coefficients, and it is possible that other choices are possible with similar
results. The particular coefficients that we use are given in Tab.~\ref{tab:PNcoeffs}. 
We note that the 3PN coefficients depend roughly monotonically on the spin of the black 
holes, but we will not attempt to infer any significance on the particular values that we obtain. 
It is important to bear in mind also that these coefficients are chosen to achieve the 
best agreement over \emph{only ten GW cycles}, and those cycles are near merger,
where the PN approximation is close to breaking down. The ``correct'' 3PN and 3.5PN
coefficients, when they are derived analytically, will be such that they will be expected
to lead to a GW phase that is physically correct at low, not high, frequencies. The
coefficients that we are using may have the reverse properties, and lead to a poor
estimate of the phase at low frequencies --- and in fact our ad-hoc coefficients may
adversely distort the phase function during the early inspiral. If anything, this will 
lead us to conclude that our waveforms should be much longer than they really 
need to be, and so our results will still provide an upper bound on the necessary
waveform length. And this is our goal. 

\begin{figure}[t]
\centering
\includegraphics[width=95mm]{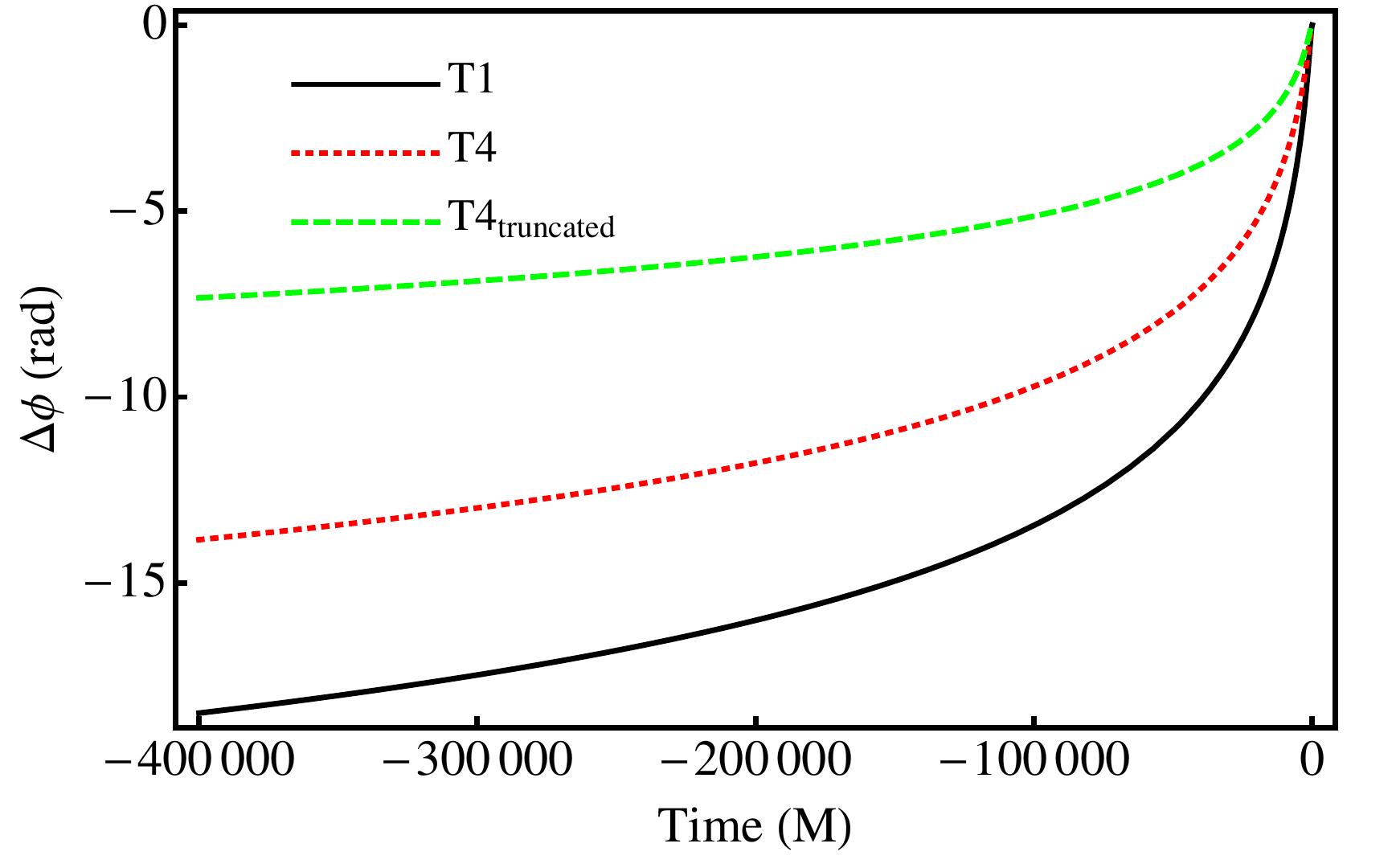}
\caption{
Phase disagreement between the optimized 3.5PN TaylorT4 approximant, and 
three other approximants: TaylorT1, T4 and T4$_{\rm truncated}$,
for the $\chi = 0.5$ case. The time $t=0$ corresponds to the frequency $M\omega
= 0.06$,
where the phases are lined up.
}
\label{fig:T1T4dphi}
\end{figure}

This point is illustrated in Fig.~\ref{fig:T1T4dphi}, which shows the accumulated
phase disagreement between the optimized 3.5PN TaylorT4 approximant, and our
three standard approximants, TaylorT1, TaylorT4, and TaylorT4$_{\rm truncated}$ 
for the $\chi = 0.5$ case. The results in~\cite{Hannam:2010ec} show that 
T4 agrees best with the NR data over the ten cycles
of comparison, T1 is the next best, and T4$_{\rm truncated}$ performs worst. 
But if we look now at Fig.~\ref{fig:T1T4dphi}, which shows the phase disagreement
during all of the early inspiral up to $M\omega = 0.06$, we get a different impression. 
Now the T4$_{\rm truncated}$ approximant performs best, while T4
is next best, and T1 is the worst. This confirms our suspicion that the relative performance
of different approximants during the last orbits of the binary may be quite different to
their performance during the earlier inspiral --- but it also tells us that if we compare
the hybrids produced by the optimized 3.5PN T4 approximant with TaylorT1 hybrids, 
then this will give us the most conservative estimate of the length requirements of
our waveforms. This is because it is the phase disagreement that dominates the 
mismatch, and Fig.~\ref{fig:T1T4dphi} suggests that the T1-T4$_{\rm 3.5PN}$ 
mismatches will be the worst.

\begin{figure}[t]
\centering
\includegraphics[width=95mm]{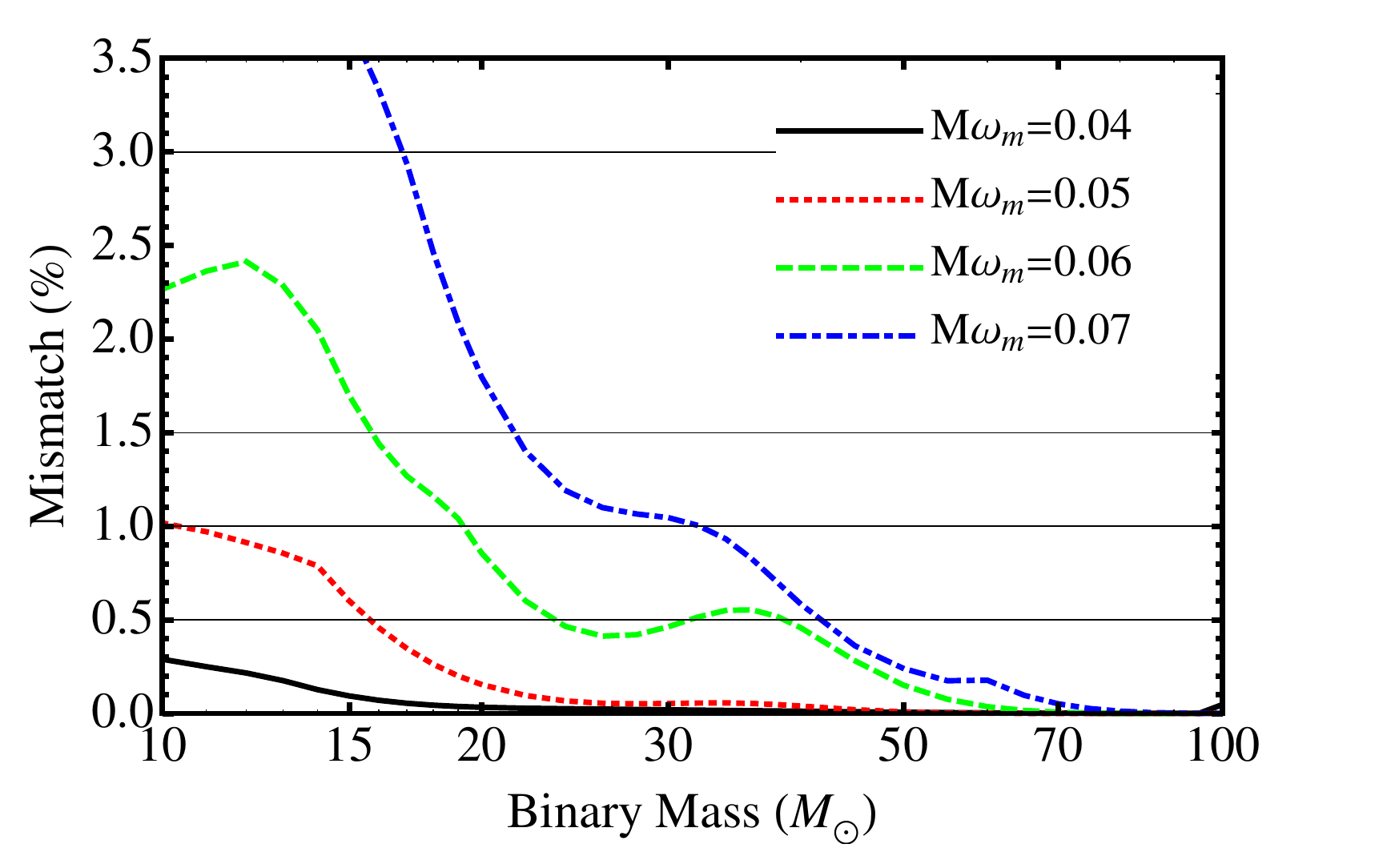}
\includegraphics[width=95mm]{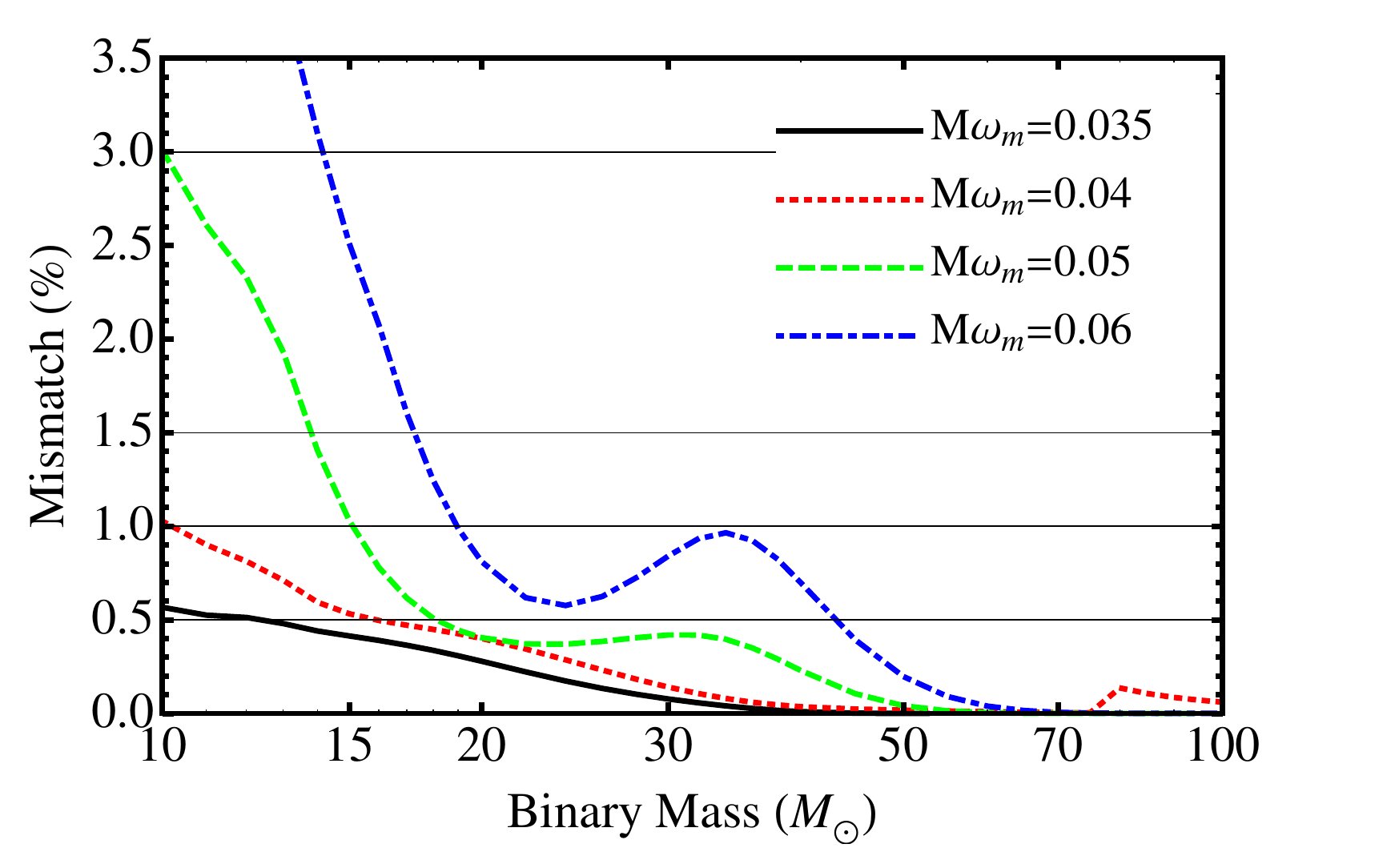}
\caption{
Mismatches between T1 and T4 hybrids for the cases $\chi = -0.5$ (upper panel) 
and $\chi = 0.5$ (lower panel). 
}
\label{fig:spinmismatches}
\end{figure}

We will focus on two cases, $\chi = -0.5$ and $\chi = 0.5$. The
mismatch plots for these cases are shown in Fig.~\ref{fig:spinmismatches}.
The most notable aspect of these two plots is the dramatic difference in the 
mismatches as a function of matching frequency between the two cases. 
For the anti-hangup case $\chi = -0.5$, a matching frequency of about 
$M\omega_m = 0.055$ appears to be sufficient to achieve mismatches below
1.5\%. This corresponds to about 10 cycles before merger, and is comparable
to what we saw in the nonspinning case. For the hangup case $\chi = 0.5$,
on the other hand, a matching frequency of between 0.04 and 0.05 is necessary.
A matching frequency of $M\omega_m = 0.04$ corresponds to 28 cycles before
merger, or 13 orbits. Some of this difference is due to the effect of spin on the 
rate of inspiral: from a given frequency, there will be more cycles before merger
in the hangup case than in the anti-hangup case. Since the hangup case requires
matching at a lower frequency, this adds yet more cycles to our estimate --- and so 
we end up with almost a factor of three difference in the number of NR cycles that are
required. 

To test the robustness of our results, we have also calculated the best-fit 
modifications to the 3PN and 3.5PN by optimizing over a smaller number of 
NR cycles. There is a large variation in the parameters as the number of 
included NR cycles is changed, but we have found that the change in the 
overall mismatch results is at a level that would not be distinguishable in the 
figures.

Tab.~\ref{tab:cycles} summarizes the number of cycles required before merger 
for the different cases, for different levels of mismatch requirement. Spins up to only 
$|\chi| = 0.5$ are used, because in higher-spin cases the deviation of the known T4 approximant 
from the NR phase is so large that the required 3PN and 3.5PN 
coefficients appear to distort the overall phase evolution too severely. We can infer
from this table, however, that for $\chi < -0.5$, simulations that include 20 GW cycles
before merger (10 orbits) should allow the construction of T1 hybrids that are within the 
3\% mismatch accuracy requirement, but for high $\chi > 0.5$ many more cycles may
be required. 

The table also indicates, assuming that we will allow a maximum error mismatch 
up to 3\%, the minimum mass that can be searched for using hybrids 
constructed from the numerical waveforms presented here. These numbers are
provided as a snapshot of what can be done with the waveforms that exist as of
the writing of this paper. In most cases we can produce hybrids that are acceptable
for searches down to 10\,$M_\odot$, but for some spinning cases we can use 
our hybrids to search down to only 15\,$M_\odot$. 
We hope that the results in this paper ultimately indicate \emph{upper bounds} on the 
the required length of numerical waveforms for GW detections. 
It should be clear from Tab.~\ref{tab:cycles} that an improvement in PN approximants 
(due for example to the calculation of higher-order spin terms) would have a 
significant effect on the NR waveform length requirements. If the accuracy of 
spinning 
approximants were improved to the level of their equal-mass nonspinning counterparts,
then numerical waveforms covering only 5-7 orbits before merger would be
sufficient for GW searches. 

On the other hand, we also see that the length requirements vary significantly between
binary configurations, and that \emph{more} NR cycles are required as we approach
the extremes of our parameter choices. We cannot determine from these results what
the length requirements would be for $q>4$ nonspinning waveforms, or even for
highly spinning waveforms with $q>1$. In this sense, the ``worst'' cases remain to be
studied.

\begin{table}
\caption{\label{tab:cycles}
Summary of mismatch calculations. The second, third and fourth columns
give the minimum number of numerical GW cycles before merger 
that ensure mismatches below 3\%, 1.5\% and
0.5\% for all masses above 10\,$M_\odot$. The last column indicates the 
lowest mass for which the numerical waveforms studied here could be used for
searches, assuming that the mismatch error can be as high as 3\%. 
}
\begin{tabular}{|| l | c | c | c | c ||}
\hline
Configuration & ${\cal M} < 3$\% & ${\cal M} < 1.5$\% & ${\cal M} < 0.5$\% &
$M_{\rm min}/M_\odot$ \\
\hline
$\chi = -0.5$   & 8.0  & 10.0 & 19.0 & 10 \\
$\chi = -0.25$ & 10.0 & 15.0 & 20.0 & 10 \\
$\chi = 0$   &  7.0  & 9.5 & 15.0 & 10 \\
$\chi = 0.25$ & 13.0 & 18.0 & 26.0 & 10 \\
$\chi = 0.5$   & 20.0 & 26.0 & 36.0 & 15 \\
$q=2$   & 8.5  & 11.5 & 25.0 & 10 \\
$q=3$   &  11.0 & 15.5 & 25.0 &  10 \\
$q=4$   &  15.0  & 21.0 & 33.0 & 10 \\
\hline
\end{tabular}
\end{table}

\section{Parameter estimation} 
\label{sec:PE}

Our focus so far has been on GW detection, and a full parameter-estimation study is
beyond the scope of this paper; indeed such a study would require a complete waveform
family, and would be more appropriately performed using phenomenological or EOB
models. However, we can make some observations about parameter estimation.

As discussed in Sec.~\ref{sec:definitions}, if for our T1 and T4 hybrids we have 
$||\delta h|| < 1$, then the two waveforms are indistinguishable. This means that
the accuracy of the estimation of the intrinsic parameters of the binary is determined by the SNR
of the signal, and not by any error in the waveforms. In other words, if the waveforms
are indistinguishable at the SNR of a given measurement, then the maximum 
parameter information can be extracted from that measurement, and is not limited
by the accuracy of the waveforms. 

It is reasonable to expect SNRs as high as 30 in Advanced detectors, and for waveforms 
to be indistinguishable at that SNR, the mismatch error must be below 0.05\% 
(see again the discussion in Sec.~\ref{sec:definitions}). It is clear even in the best
case (equal-mass nonspinning) that our hybrids meet this criteria only for binary 
masses higher than 20\,$M_\odot$. To achieve such a low mismatch down to 
10\,$M_\odot$ would require numerical waveforms  far longer than any that
have yet been produced. We have also seen that the mismatch in
the hybrids due to artifacts from the hybridization process alone are at 0.03\%,
and so producing hybrids that are indistinguishable for parameter estimation up
to an SNR of 30 is a challenge irrespective of the problems of PN errors and
waveform length. 

However, even though the hybrids we have presented may not be indistinguishable 
at the potential SNRs of Advanced detectors, they can still be used to estimate the 
parameters of a binary, and the question remains what the accuracy of the parameter
estimation can in principle be, and whether this accuracy is sufficient for 
GW astronomy applications in the near future. 

\begin{figure}[t]
\centering
\includegraphics[width=85mm]{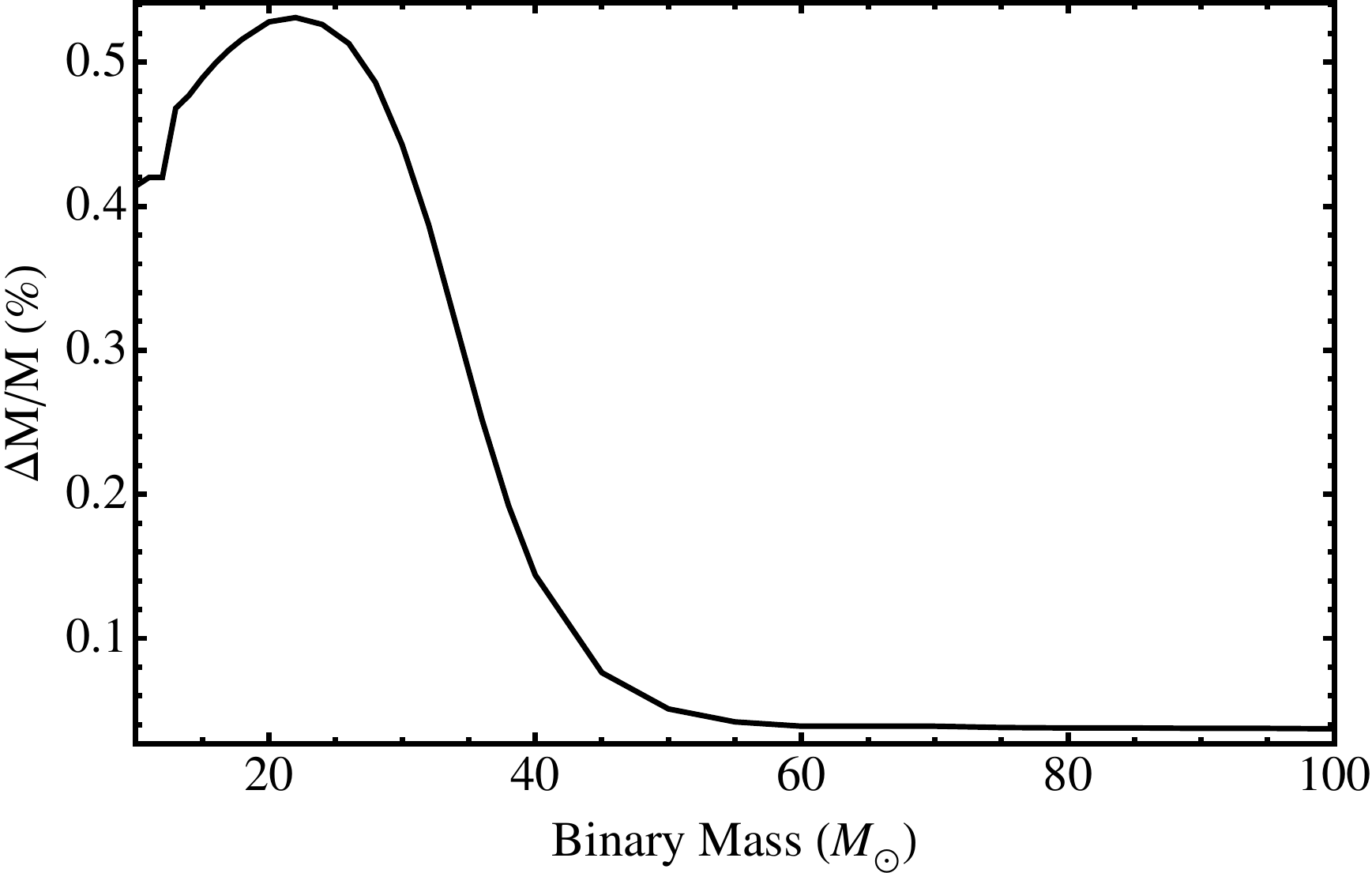}
\caption{
Bias in mass estimation for the $\chi = 0.5$ case, based on a comparison
of T1 and T4 hybrids.
}
\label{fig:T1T4bias}
\end{figure}

To give an indication of what these errors might be, we can calculate the mass bias
from our mismatch calculations. For the least accurate case we have considered 
here, $\chi = 0.5$, with a hybrid matching frequency of $M\omega_m = 0.05$, 
the worst mass bias is 0.54\% for $M=22\,M_\odot$. The results for masses up to 
$100\,M_\odot$ are shown in Fig.~\ref{fig:T1T4bias}. We note for comparison that 
the results in Ref.~\cite{Buonanno:2009zt} suggest that the 
PN waveforms that can be used for searches for binaries with $M < 10\,M_\odot$ have 
relatively high mass biases, around 20\% for $M = 10\,M_\odot$. In addition, 
the results in Ref.~\cite{Bose:2009} suggest that the mass bias from using 35PN TaylorT1 
inspiral templates to detect phenomenological waveforms is around 10\%.
These results of course also suggest that hybrids that are accurate enough for parameter 
estimation are also needed for binaries \emph{below} $10\,M_\odot$. 

We emphasize that the true bias in the mass, when optimizing over all of the intrinsic
and extrinsic parameters, may differ from that indicated in Fig.~\ref{fig:T1T4bias}. 
While these results suggest that mass estimation 
errors will be very small with these waveforms, we defer a detailed analysis of 
parameter estimation errors to future work.

\section{Conclusions}
\label{sec:conclusion}

In constructing hybrid PN+NR black-hole-binary waveforms there are three evident
sources of error: the error in the numerical waveforms, the error in the PN waveforms,
and the error introduced in the hybridization process. The error measure relevant for
GW \emph{detection} is the mismatch error, and in the context of current GW searches 
this error is required to be less than 3\%. We know from previous 
work~\cite{Hannam:2009hh,Hannam:2010ec} that the mismatch error in the numerical 
waveforms is within this requirement by several orders of magnitude. 
We have shown in this paper that the mismatch error
due to the hybridization process is also very low, at $\sim 0.03$\%. The mismatch error 
due to the PN contribution to the hybrids, however, can be much larger, and 
dominates the error budget in all hybrids that can be produced with current PN and 
NR results. 

Motivated by these observations, we have attempted to address the question of 
how many NR cycles (i.e., how few PN cycles) must be used in a hybrid in order
for the hybrid to be sufficiently accurate for GW detection purposes. Our approach,
which we justify in detail in Sec.~\ref{sec:length}, is to construct TaylorT4+NR hybrids,
and to treat these as the ``true'' GW signal (NR$_{\rm L}$), and then to calculate the 
mismatch between these and hybrids of TaylorT1+NR$_{\rm L}$. The mismatch is
optimized with respect to the total mass of the binary, which we argue in 
Sec.~\ref{sec:lengthUM} is close to the mismatch resulting from a calculation of 
the fitting factor, which is the relevant quantity for GW detection. This allows us
to make the most conclusive statements possible with currently available numerical
simulations. 

In the equal-mass
nonspinning case we find that very few NR cycles are necessary; five cycles (two orbits) 
are sufficient to be within the 3\% mismatch requirement. Phase comparisons between 
PN and NR results suggest that the relative accuracy between TaylorT1 and TaylorT4
does not change significantly as the mass ratio is increased to 
$q=4$~\cite{Hannam:2010ec}, so we might expect that the length requirements are 
roughly the same for larger mass ratios. On the contrary, we find that the length
requirements increase with the mass ratio, and for $q=4$ 15 cycles (seven orbits)
are necessary. We conclude that (1) we cannot infer NR-waveform length 
requirements directly from PN-NR phase comparisons, and (2) at higher mass
ratios $q>4$ we will require yet longer NR waveforms, but more extensive studies
will be required to determine how many.  

Due to the greater PN errors for spinning binaries, more cycles are generally needed 
than in the nonspinning case. Our results are summarized in Tab.~\ref{tab:cycles}. 
We have considered only \emph{equal-mass} spinning cases, but we can conclude 
from these results that for \emph{unequal-mass} spinning binaries, more NR cycles will
be needed than in any of the cases we have considered here.

A conservative summary of our results would be that in the cases we have considered, 
simulations of 10 orbits before merger should be sufficient to produce hybrids that are
accurate enough for GW detection purposes. The resulting hybrids will be indistinguishable
for SNRs in Advanced detectors of less than 30 for masses above $20-30\,M_\odot$. 
The important caveat to 
our results is that they apply only to the ``cheapest'' cases. Nonspinning binaries with 
$q>4$ and unequal-mass spinning binaries will require more NR cycles, and we have 
not considered any cases with precessing spins. 

These conclusions depend strongly on the current state of the art. The advent of more 
accurate PN  results (for example higher order spin contributions) could reduce these 
length requirements, as might a more robust quantification of the errors in PN methods. 
We have also not considered EOB results, for which the errors may be far lower, with a 
corresponding drop in the NR-waveform length requirements. 

In addition to being relevant to the construction of hybrids for analytic waveform models,
our results also have direct bearing on two current efforts in the NR, data-analysis and
analytical modeling communities. 

In the second stage of the NINJA project~\cite{ninja-wiki} hybrid 
waveforms are being injected into detector noise to test search pipelines. The hybrids
are constructed with matching frequencies in the range we have considered, but
in most cases the matching will be around $M\omega = 0.07$. In general our 
results suggest that these hybrids will have a mismatch error of greater than 3\% for
masses lower than $20-30\,M_\odot$. 

The complementary NR-AR project~\cite{ninja-wiki} represents a community-wide effort 
to produce a large number of NR waveforms to calibrate analytic models. 
The nominal requirement for these waveforms is that they include 20 GW cycles 
(10 orbits) before merger. From the 
point of view of constructing hybrid waveforms for GW detection with typical PN 
approximants, our results suggest that these waveforms will be long enough along the 
branches of the parameter space that we have considered, but possibly not for
high-mass-ratio cases, some unequal-mass spinning cases, and perhaps also cases
with precession. 

We have presented a general method to assess NR waveform length requirements, 
but our study only begins to address this question. Further work is needed to develop
an easy-to-implement and robust method to estimate these requirements for 
configurations that have \emph{not yet} been simulated in NR codes; work in this
area is already underway~\cite{Ohme:2010aa}. It would be useful
to make estimates of waveform length requirements for the construction of analytic
models, both phenomenological and EOB; to extend this study to possible future
detectors, like LISA~\cite{Shaddock:2009za} and the Einstein 
Telescope~\cite{Punturo:2010zza}; and, finally, much more work is 
required to understand the length requirements for parameter estimation, and to 
balance the needs of GW astronomy with the computational cost of numerical 
simulations.

\section*{Acknowledgments}

We thank Steve Fairhurst and Harald Pfeiffer for useful discussions, and 
Sukanta Bose and Doreen M\"uller for helpful comments on the manuscript.
M. Hannam was supported by an FWF Lise Meitner fellowship (M1178-N16).
S. Husa was supported by
grant FPA-2007-60220 from the Spanish Ministry of Science,
the Spanish MICINN’s Consolider-Ingenio 2010 Programme under grant 
MultiDark CSD2009-00064, and DAAD grant D/07/13385.
F. Ohme thanks the IMPRS for Gravitational Wave Astronomy and the DLR (Deutsches
Zentrum f\"ur Luft-und Raumfahrt) for support.
PA was supported in part by NSF grants PHY-0653653 and 
PHY-0601459, and the David and Barbara Groce Fund at Caltech.
{\tt BAM} simulations were carried out at LRZ Munich, ICHEC Dublin,
the Vienna Scientific Cluster (VSC), at MareNostrum at Barcelona 
Supercomputing Center -- Centro Nacional de Supercomputaci\'on 
(Spanish National Supercomputing Center), and CESGA, Santiago the Compostela.

\bibliography{NonPrecessing}

\end{document}